\documentclass[traditabstract]{aa}
\usepackage[utf8]{inputenc}
\usepackage{threeparttable,pifont}
\usepackage{natbib}
\usepackage{color}
\usepackage{multicol}
\usepackage{amsmath,amssymb,url}
\usepackage{graphicx}
\usepackage{xcolor}
\usepackage{txfonts,longtable,dcolumn,verbatim}
\bibpunct{(}{)}{;}{a}{}{,}
\usepackage{array} 
\usepackage[normalem]{ulem}
\usepackage{lscape}

\usepackage[squaren,textstyle]{SIunits}

\usepackage{hyperref}
\hypersetup{
    unicode=true,           
    pdftoolbar=true,        
    pdfmenubar=true,        
    pdffitwindow=true,      
    pdftitle={The V-Band Photometry of Asteroids from ASAS-SN},
    pdfauthor={Hanus et al.},
    pdfsubject={Planetary Science},
    pdfkeywords={},         
    pdfnewwindow=true,      
    colorlinks=true,        
    linkcolor=gray,         
    citecolor=blue,         
    filecolor=gray,         
    urlcolor=gray           
}

\usepackage{xspace}

\begin{document}

\title{$V$-band photometry of asteroids from ASAS-SN}
\subtitle{Finding asteroids with slow spin
    \thanks{Tables A.1-4 are  available  at  the  CDS  via  anonymous   ftp  to \url{http://cdsarc.u-strasbg.fr/} or via \url{http://cdsarc.u-strasbg.fr/viz-bin/qcat?J/A+A/xxx/Axxx}}}

\titlerunning{$V$-Band Photometry of Asteroids from ASAS-SN}

\author{  
    J.~Hanu{\v s}\inst{\ref{auuk}}       \and 
    O.~Pejcha\inst{\ref{teorka}}           \and 
    B.~J.~Shappee\inst{\ref{hawaii}}          \and 
    C.~S.~Kochanek\inst{\ref{osu},\ref{ccapp}} \and
    K.~Z.~Stanek\inst{\ref{osu},\ref{ccapp}}  \and
    T.~W.-S.~Holoien\inst{\ref{carnegie},}\thanks{NHFP Einstein Fellow}
} 

\institute{
     Institute of Astronomy, Faculty of Mathematics and Physics, Charles University, V~Hole{\v s}ovi{\v c}k{\'a}ch 2, 18000 Prague, Czech Republic\label{auuk}
     \and 
     Institute of Theoretical Physics, Faculty of Mathematics and Physics, Charles University, V~Hole{\v s}ovi{\v c}k{\'a}ch 2, 18000 Prague, Czech Republic
     \label{teorka}
     \and 
     Institute for Astronomy, University of Hawai'i, 2680 Woodlawn Drive, Honolulu, HI 96822, USA
     \label{hawaii}
     \and
     Department of Astronomy, The Ohio State University, 140 West 18th Avenue, Columbus, OH 43210, USA
     \label{osu}
     \and
     Center for Cosmology and Astroparticle Physics, The Ohio State University, 191 W. Woodruff Avenue, Columbus, OH 43210, USA\label{ccapp}
     \and
     The Observatories of the Carnegie Institution for Science, 813 Santa Barbara St., Pasadena, CA 91101, USA
     \label{carnegie}
}

   \date{Received x-x-2021 / Accepted x-x-2021}
 
\abstract
{We present $V$-band photometry of the 20,000 brightest asteroids using data from the All-Sky Automated Survey for Supernovae (ASAS-SN) between 2012 and 2018. We were able to apply the convex inversion method to more than 5,000 asteroids with more than 60 good measurements in order to derive their sidereal rotation periods, spin axis orientations, and shape models. We derive unique spin state and shape solutions for 760 asteroids, including 163 new determinations. This corresponds to a success rate of about 15\%, which is significantly higher than the success rate previously achieved using photometry from surveys. We derive the first sidereal rotation periods for additional 69 asteroids. We find good agreement in spin periods and pole orientations for objects with prior solutions. We obtain a statistical sample of asteroid physical properties that is sufficient for the detection of several previously known trends, such as the underrepresentation of slow rotators in current databases, and the anisotropic distribution of spin orientations driven by the nongravitational forces. We also investigate the dependence of spin orientations on the rotation period. Since 2018, ASAS-SN has been observing the sky with higher cadence and a deeper limiting magnitude, which will lead to many more new solutions in just a few years.
} 


\maketitle

\section{Introduction}\label{sec:introduction}

A large number of sky surveys have monitored the sky with various depths, cadences, coverage areas, and scientific goals. Examples include the All-Sky Automated Survey \citep[ASAS;][]{pojmanski2002}, the Optical Gravitational Lensing Experiment \citep[OGLE;][]{udalski2003}, the Northern Sky Variability Survey \citep[NSVS;][]{wozniak2004}, MACHO \citep{alcock1997}, EROS \citep{palanque1998}, the Catalina Sky Survey \citep[CSS,][]{Larson2003}, the Asteroid Terrestrial-impact Last Alert System \citep[ATLAS;][]{tonry2018}, \textit{Gaia} \citep{prusti2016}, the Zwicky Transient Facility \citep[ZTF;][]{bellm2019}, \textit{Kepler} \citep{Borucki2010}, \textit{TESS} \citep{Ricker2015}, and PanSTARRS \citep{Chambers2016}. Although several surveys were designed specifically to find asteroids, most of them are not primarily dedicated to the study of Solar System bodies, but could be used to do so with some effort \citep[e.g.,][]{Durech2020,Pal2020,Szabo2016,Spoto2018}.

Long-term calibrated photometry from modern time-domain surveys both from the ground and from space has been an important tool for the physical characterization of Solar System bodies. For example, the Sloan Digital Sky Survey data revealed a bimodal distribution of the broadband colors of main-belt asteroids caused by differences between rocky and carbonaceous surface compositions \citep{Ivezic2001}. Time series data from \textit{K2} \citep{Szabo2016} and \textit{TESS} \citep{Pal2020} were used to determine rotation periods for a large number of asteroids.

The information content in the light curves is not limited to the rotation periods only. \citet{Durech2010} presented a database of asteroid rotation periods, pole orientations, and three-dimensional (3D) convex shapes determined using the photometry inversion technique of \citet{Kaasalainen2001a} and \citet{Kaasalainen2001b}. There are currently (as of February 2021) 5,715 shape models for 3,303 asteroids,\footnote{\url{https://astro.troja.mff.cuni.cz/projects/damit/}} and the number of solutions is continuously expanding as new data become available. For example, \citet{Durech2018d} and \citet{Durech2020} determined 3D convex shape models and rotation pole orientations for 129 and 1,800 asteroids using Gaia DR2 and ATLAS photometry, respectively.

Understanding the distribution of spin vectors is important for constraining the evolution of asteroid rotational states. \citet{Hanus2011} and \citet{Hanus2013a} identified the signatures of the thermal Yarkovsky--O'Keefe--Radzievskii--Paddack effect \citep[YORP,][]{Bottke2006}, which pushes the spin orientation of smaller asteroids to align perpendicularly to their orbital planes. Even better constraints on individual asteroid properties can be obtained by combining shape modeling with infrared photometry \citep{Hanus2015a,Hanus2018b,Durech2017a}, occultations of stars \citep{Durech2011}, adaptive optics imaging \citep{Hanus2013b,Hanus2017b, Viikinkoski2017, Viikinkoski2018, Vernazza2018}, and interplanetary mission flybys \citep{Preusker2016a, Preusker2016b, Preusker2017, Watanabe2019}.

\begin{figure}
    \centering
    \includegraphics[width=0.48\textwidth]{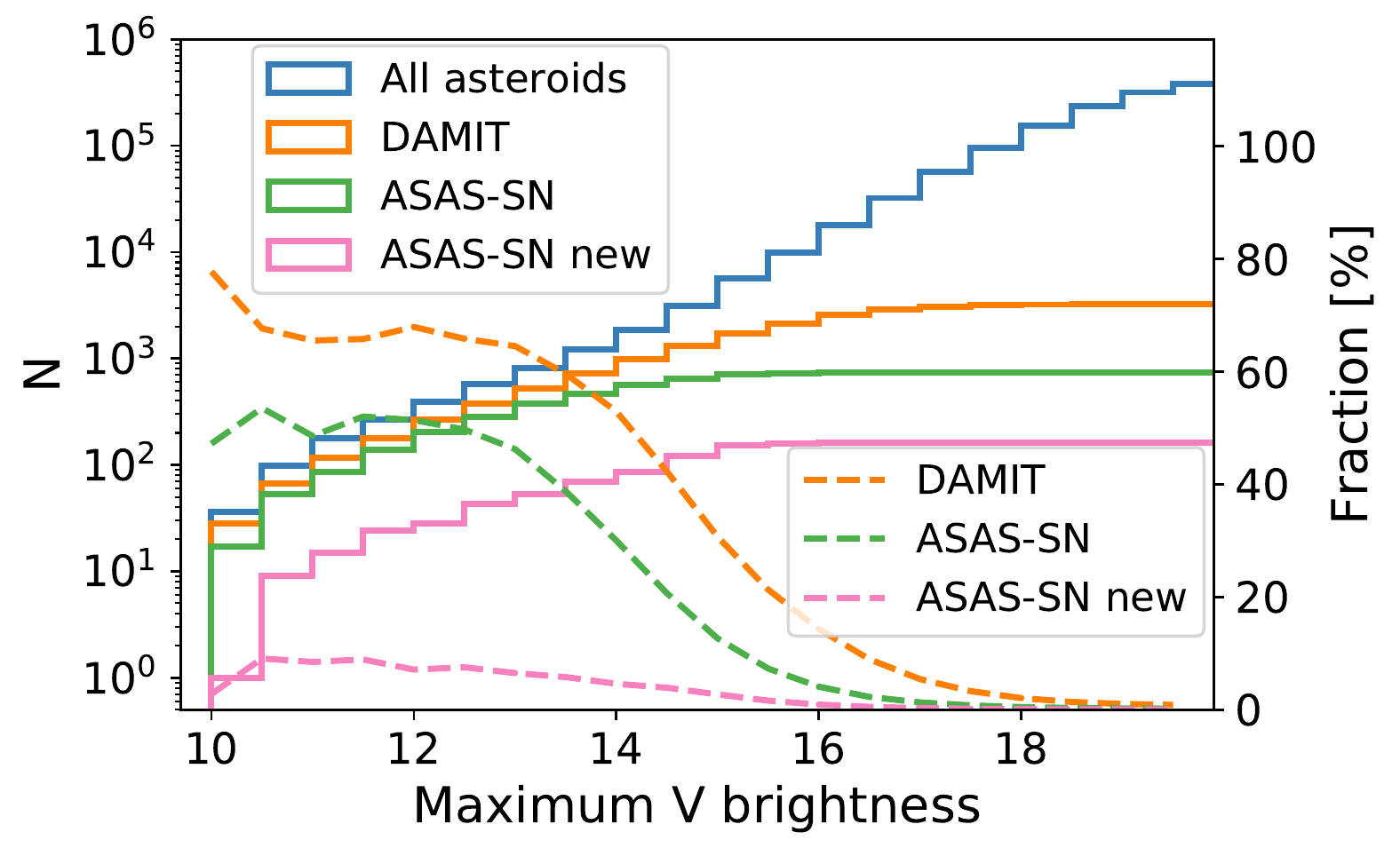}\\
    \caption{Distributions of asteroids and available shape models as a function of maximum $V$-band magnitude. The solid lines show all known asteroids (blue), shape solutions available in the DAMIT database (orange), shape solutions obtained from ASAS-SN data in this work (green), and the asteroids with new shape models from this work (pink). The dashed lines show the percentage of all asteroids that have shape solutions.}
    \label{fig:bright_ast}
\end{figure}

Bright asteroids are often the most suitable targets for detailed studies, but the currently available data do not provide a complete picture of the asteroid population. In Figure~\ref{fig:bright_ast} we show the fraction of asteroids that have shape models as a function of their maximum $V$-band magnitude. We see that the completeness is about 70\% at the bright end, and begins to decrease for $V>14$\,mag. More observations, even of bright asteroids, are needed.

In this paper we present $V$-band photometry of asteroids obtained with the All-Sky Automated Survey for Supernovae \citep[ASAS-SN,][]{Shappee2014,Kochanek2017} between the years 2012 and 2018. During this period ASAS-SN monitored the sky  with a cadence of 2--3 days to a depth of $V \sim 17$\,mag using two units in Chile and Hawaii. Although the primary goal of ASAS-SN is the hunt for supernovae \citep{Holoien2017} and other transients such as tidal disruption events \citep{Holoien2014} or novae \citep{Aydi2020}, it has been used much more broadly to, for example, characterize stellar variability \citep{Jayasinghe2018,Jayasinghe2019,Jayasinghe2020b,Pawlak2019,Bredall2020} or even to discover comets \citep[ASASSN1 or C/2017 O1 and ASASSN2 or C/2018 N2,][]{Prieto2017,Brinkman2020,vanBuitenen2018}. In this paper we use the calibrated ASAS-SN disk-integrated photometry for the physical characterization of asteroids, in terms of their sidereal rotation periods, spin axis orientation, and 3D shape models.

The structure of our paper is as follows. First, we describe our ASAS-SN data processing pipeline in Sect.~\ref{sec:Vband}. Then, we outline the light curve inversion scheme, apply it to ASAS-SN photometry, and derive physical properties for several hundred asteroids in Sect.~\ref{sec:results}. We discuss our results and derived physical properties in Sect.~\ref{sec:discussion}, and conclude our work in Sect.~\ref{sec:conclusions}.

\section{ASAS-SN V-band photometry}\label{sec:Vband}

\begin{figure}
\begin{center}
\resizebox{1.0\hsize}{!}{\includegraphics{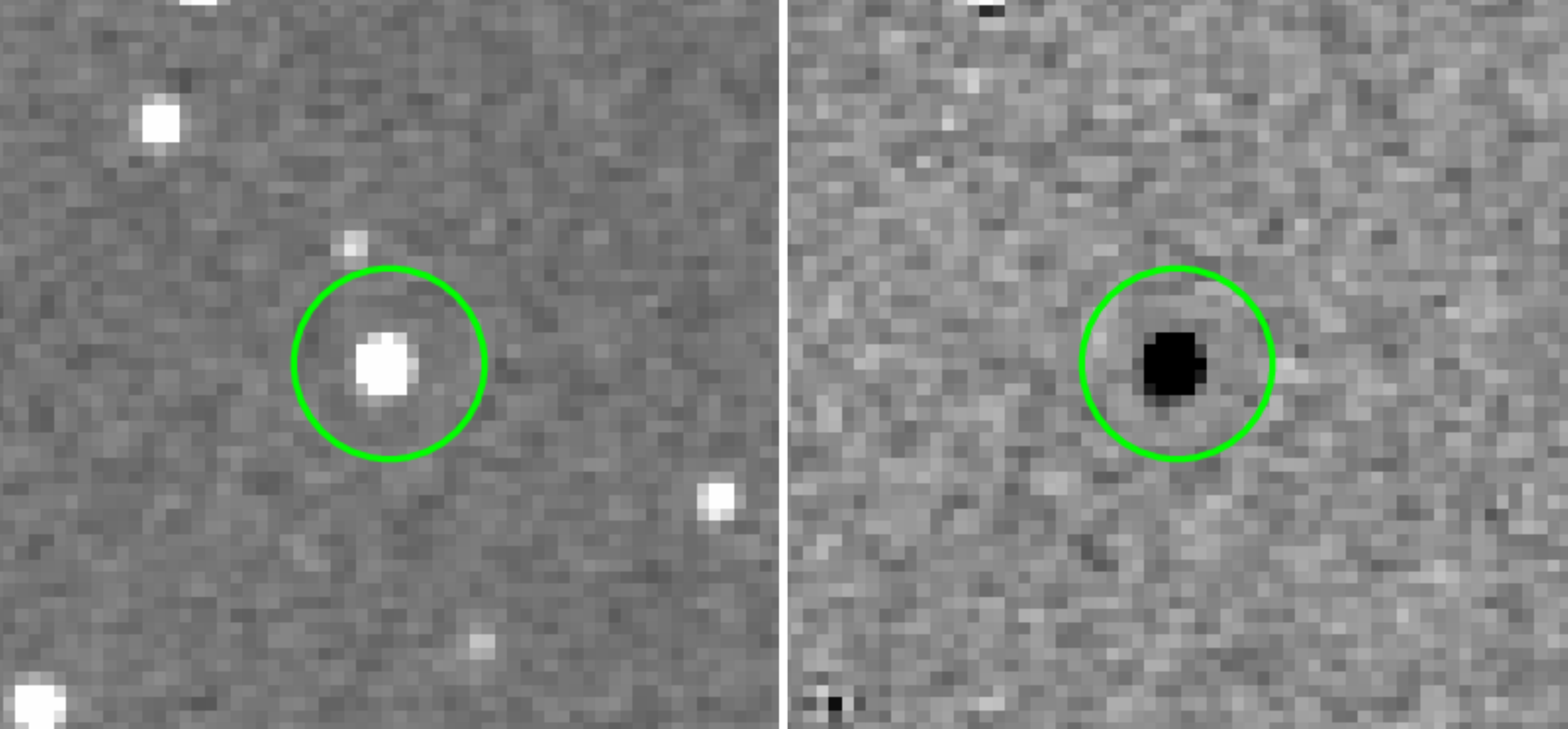}}\\
\end{center}
\caption{\label{fig:stamps}Example of typical images obtained by ASAS-SN. \textit{Left}: Image containing asteroid (130)~Elektra with an exposure time of 90 s. \textit{Right}: Corresponding subtracted image (reference image minus observation). The circle of 1 arcmin in radius is centered on the predicted coordinates of Elektra. }
\end{figure}

\begin{figure*}
\begin{center}
\resizebox{1.0\hsize}{!}{\includegraphics{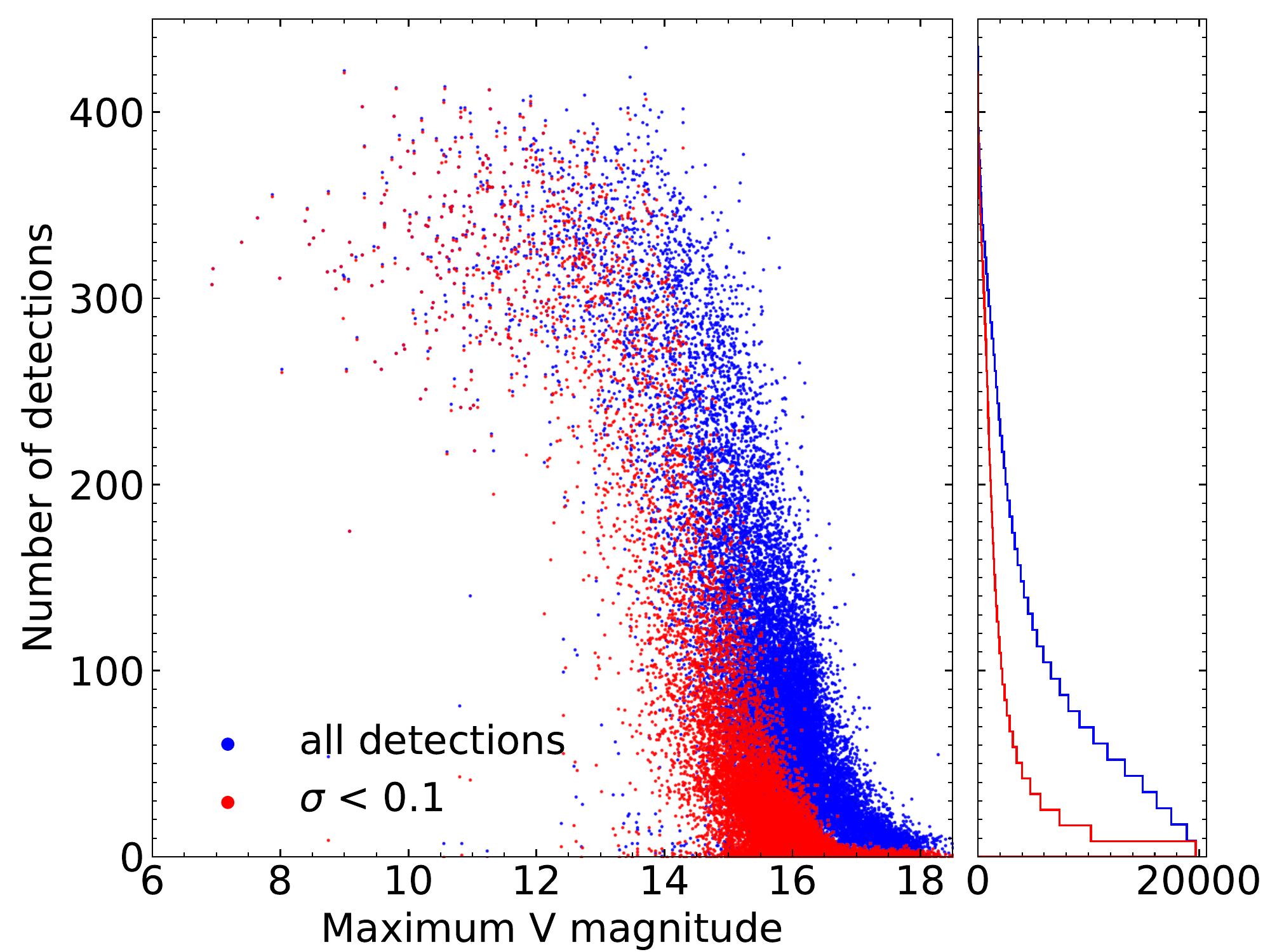}\includegraphics{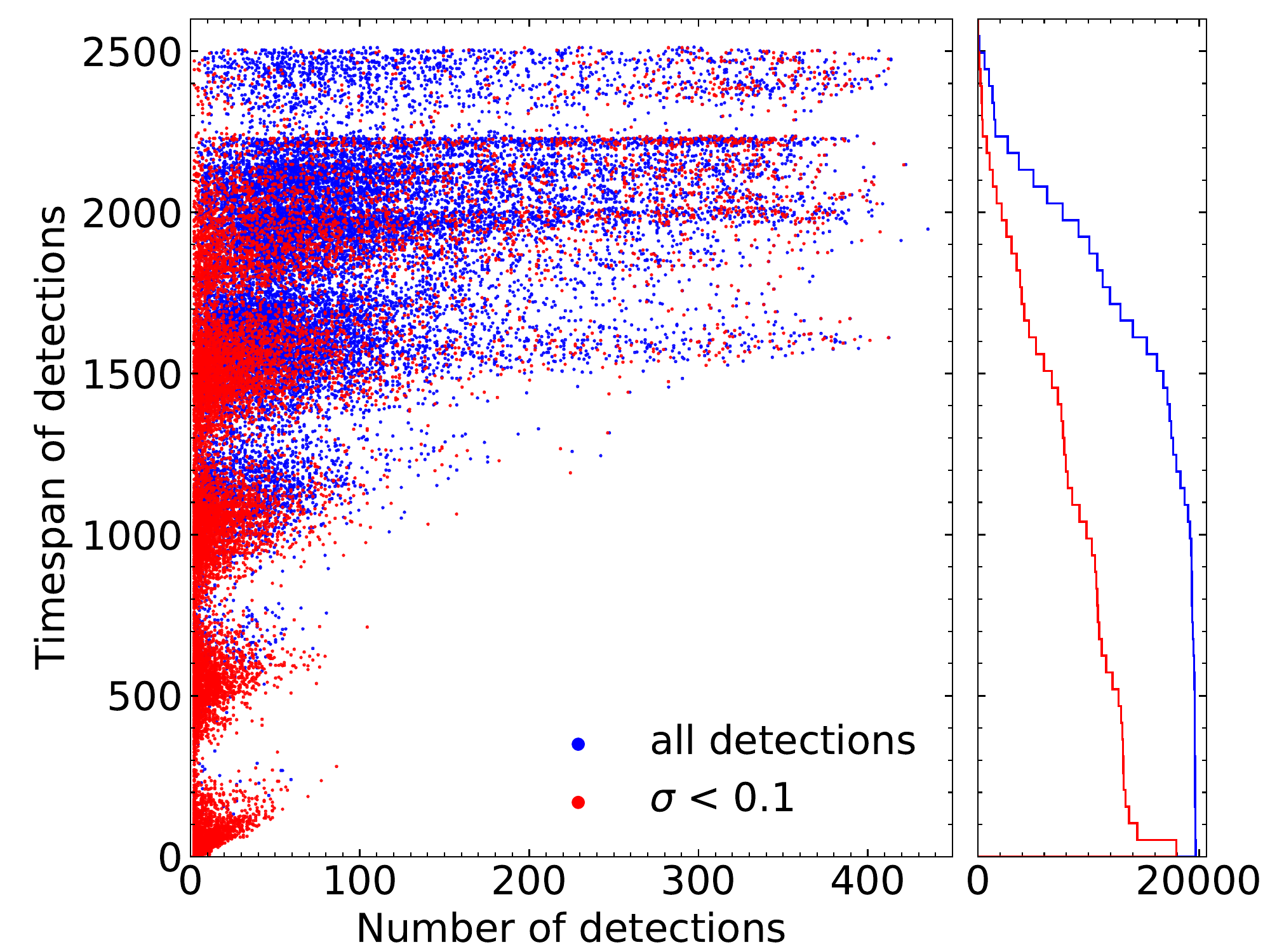}}\\
\end{center}
\caption{\label{fig:asasStat}Basic statistics of the ASAS-SN data. \textit{Left}: Number of individual detections for each asteroid as a function of the maximum $V$-band magnitude predicted by Miriade. \textit{Right}: Timespan of the observations in days as a function of the number of detections. Asteroids with average photometric uncertainties $\sigma<0.1$\,mag are shown in red.}
\end{figure*}

\begin{figure*}
\begin{center}
\resizebox{0.8\hsize}{!}{\includegraphics{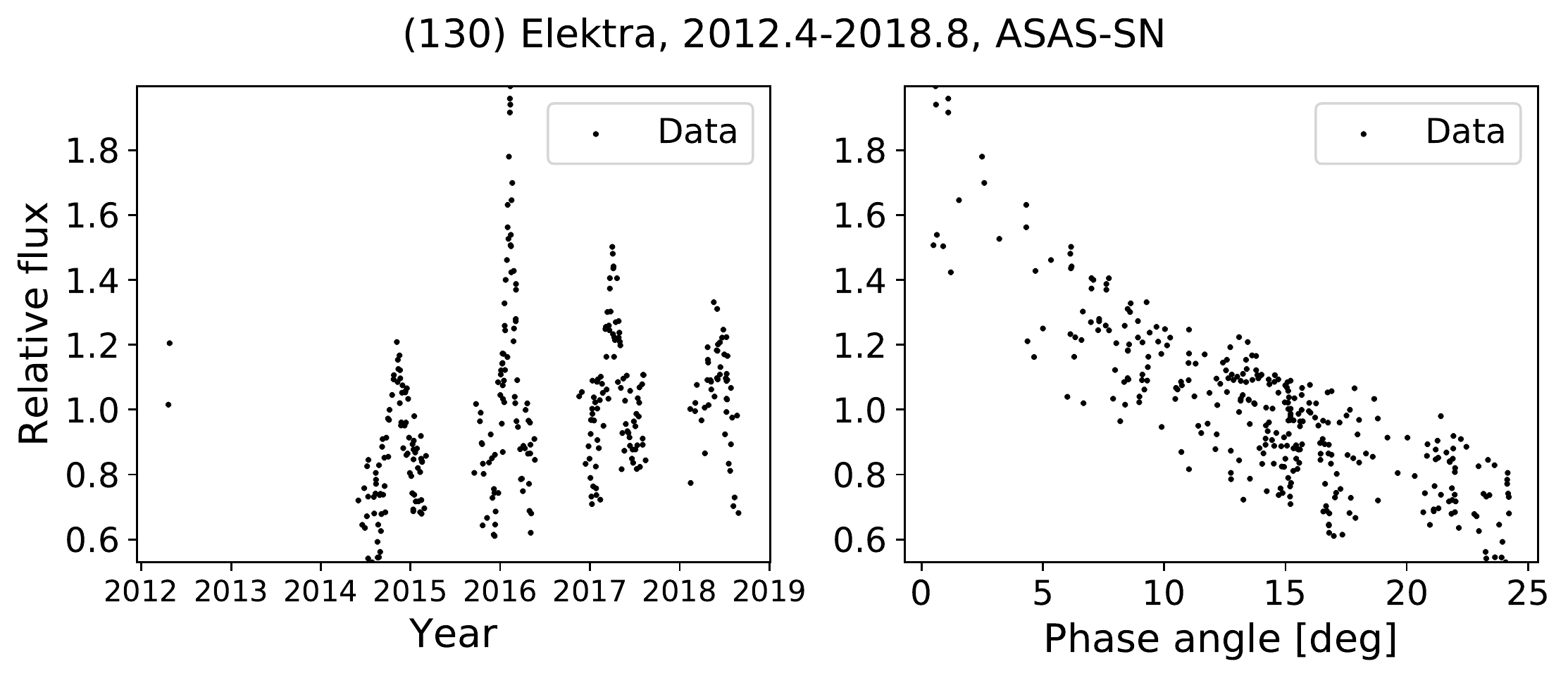}}\\
\resizebox{0.8\hsize}{!}{\includegraphics{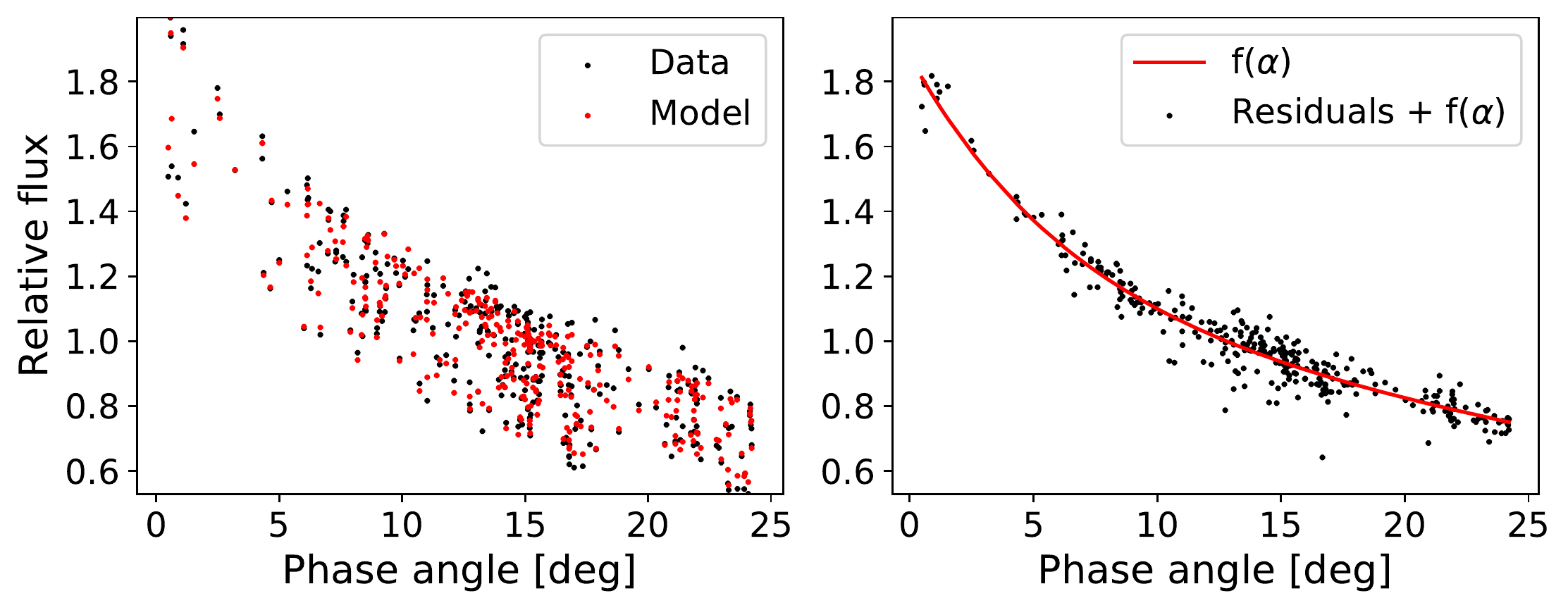}}\\
\end{center}
\caption{Typical ASAS-SN V-band photometry. \label{fig:130data}\textit{Top row}: ASAS-SN normalized $V$-band photometry of asteroid (130)~Elektra as a function of time (left) and phase angle $\alpha$ (right). \textit{Bottom row}: Comparison between the observed data and the model for asteroid (130)~Elektra as a function of phase angle $\alpha$ (left) and the fit residuals with respect to the phase function $f(\alpha)$ (right). The actual H$_{V}$ values are not relevant for our study, and therefore the relative flux is used instead.}
\end{figure*}

During 2012-2018 ASAS-SN consisted of two units in Chile and
Hawaii. Each unit consisted of a robotic mount with four 14cm
telescopes each with a $4.5^\circ \times 4.5^\circ$ field of
view.  The detectors were $2048^2$ cooled, back-illuminated
CCDs with $8\farcs0$ pixels, and the image FWHM was roughly
2~pixels. During this period, the survey used a $V$ filter with a
typical limiting magnitude of $V\sim 17$~mag under good sky
conditions. This dataset is the focus of the paper. Since
2018 ASAS-SN has expanded to five units with new mounts in
Chile, Texas, and South Africa, and switched to the Sloan $g$ filter
with a limiting magnitude of $g\sim 18.5$~mag. These observations
are continuing and will be the subject of a future work

Each ASAS-SN epoch consists of three (10 pixel) dithered images each
with an exposure time of 90~s and roughly 15~s between
exposures. The photometry is based on image
subtraction followed by aperture photometry on the difference
image. In image subtraction a reference image constructed by
combining large numbers of high quality images is scaled in flux and
PSF structure to match the new data, and is then subtracted to
leave an image consisting only of the changes from the
reference image \citep{Alard1998, Alard2000}, see an example in Fig.~\ref{fig:stamps}. Because the reference image
is built from a large number of images, its contribution
to the statistical noise is negligible. Fields are observed
to optimize the discovery of supernova-like transients based
on the age and quality of the last observation. A particular field is visited only once per night. However, there is no fixed temporal spacing \citep[see Fig.~2 in][]{Pawlak2019}, and in searches for variable stars, there is no difficulty with measuring periods of $\sim 1$~hour \citep{Jayasinghe2020a, Jayasinghe2020b}. Only simple multiples of the diurnal period have
significant power in the typical window function.

We performed the absolute calibration using the AAVSO
Photometric All-Sky Survey \citep[APASS;][]{Henden2015} catalog.
Ephemeris tables from Miriade accurately predict the location
of asteroids in the ASAS-SN data. We selected a region around
the predicted position, centered on the emission from the asteroid,
and then measured the flux with aperture photometry.

Given the limiting magnitude of $\sim$17, we decided to extract data for the first 10,000 numbered asteroids plus the remaining asteroids predicted to become brighter than $V=16.5$ mag. In total, our initial dataset involves $\sim$20,000 asteroid light curves. We show the basic statistics in Fig.~\ref{fig:asasStat}. We have at least 10 {flux measurements} for $\sim$19,000 asteroids, at least 100 for $\sim$7,000 asteroids, and at least 200 for $\sim$1,000 asteroids. The average photometric accuracy is $<0.1$\,mag for about one-third of the objects. About 80\% of the asteroids have data spanning more than 1,500 days (4.1 years) corresponding to 3--5 apparitions for each asteroid. Having data from at least three different apparitions (i.e., different viewing geometries) is crucial for successful shape modeling by the light curve inversion technique \citep{Kaasalainen2001b}. 

Our data contain a significant fraction of clear outliers. To identify them, we use Miriade\footnote{\url{http://vo.imcce.fr/webservices/miriade/}} to obtain the predicted V band magnitude for each epoch. We remove points for which the extracted magnitude differs from the predicted by more than 0.7 magnitude. We selected this threshold because the predicted magnitude reflects the light curve average, while the observed magnitude can correspond to any part of the light curve. Light curve amplitudes (peak to peak) are rarely larger than one magnitude, so our threshold should be sufficiently generous to avoid rejecting real data even for asteroids with extreme amplitudes.

Our data processing follows the standard approach applied to sky survey data \citep[e.g.,][]{Hanus2011,Hanus2013a,Durech2016,Durech2018d,Durech2020}.  
First, we obtain the asterocentric ecliptic coordinates of the Sun and the Earth using Miriade. Having the distances of the asteroid from the Earth, we correct the epochs for light travel times. Then, we transform the magnitudes to intensities, defining a magnitude of 15 to correspond to an intensity of unity for later computational convenience. Next, we normalize the flux intensities to distance of 1 au (astronomical unit) from the Earth and the Sun and average them to unity. We fit the data with a standard phase function \citep{Kaasalainen2001b}
\begin{equation}\label{eq:phase}
    f(\alpha)= A_0 \exp{\frac{-\alpha}{D}} - k\alpha + 1,
\end{equation}
where $\alpha$ is the phase angle and $A_0, D, k$ are free parameters, and then remove any remaining outliers using sigma clipping. We use a clipping threshold of 2.5$\sigma$ based on our empirical experience \citep[also see][]{Durech2020}. In cases with a small number of data points ($\lesssim$30) or a poor sampling of low phase angles ($\alpha<10^\circ$) the fitting sometimes fails. For these cases we only use the linear part of the phase function for outlier rejection (i.e., we set $a=0$). As a final step, we keep only measurements with photometric accuracy better than 0.15 mag. This threshold is a reasonable compromise between including too much data with poor photometric quality and removing too many useful measurements. Typical ASAS-SN $V$-band data are illustrated in Fig.~\ref{fig:130data} for the main-belt asteroid (130)~Elektra.

These steps rejected 30\% of the data points on average. The losses were small for bright asteroids ($<$10\% rejected for $V<15.5$\,mag), but significant for the fainter asteroids (often more than 40--50\% for $V>16$\,mag). Our catalog of ASAS-SN $V$-band photometry for 19,402 asteroids is presented in Table~\ref{tab:data}. For each photometric data point we provide the Julian date, $V$-band magnitude with the uncertainty, and flags indicating whether the data point was used in the light curve inversion.

For the purpose of the shape modeling by the convex inversion algorithm (Sect.~\ref{sec:results}), calibrated photometry normalized to unity is sufficient because we only need relative changes of the brightness. The fit of the phase function by a semi-empirical model (Eq.~\ref{eq:phase}) is part of the convex inversion. Therefore, the possible prior conversion to the more common absolute magnitudes H$_{V}$ was not needed. Anyone interested in H$_{V}$ magnitudes can use our catalog of $V$-band magnitudes and can do the analysis of the phase function independently.

\section{Determination of physical parameters by photometry inversion}\label{sec:results}

In this section we   describe the light curve inversion scheme we apply to the ASAS-SN $V$-band data in Sect.~\ref{sec:inv} and the threshold we use to evaluate the uniqueness of the solutions in Sect.~\ref{sec:threshold}. Finally, we derive new spin state and shape solutions in Sect. \ref{sec:individual}.

\subsection{Inversion scheme}\label{sec:inv}

\begin{figure*}
\begin{center}
\resizebox{1.0\hsize}{!}{\includegraphics{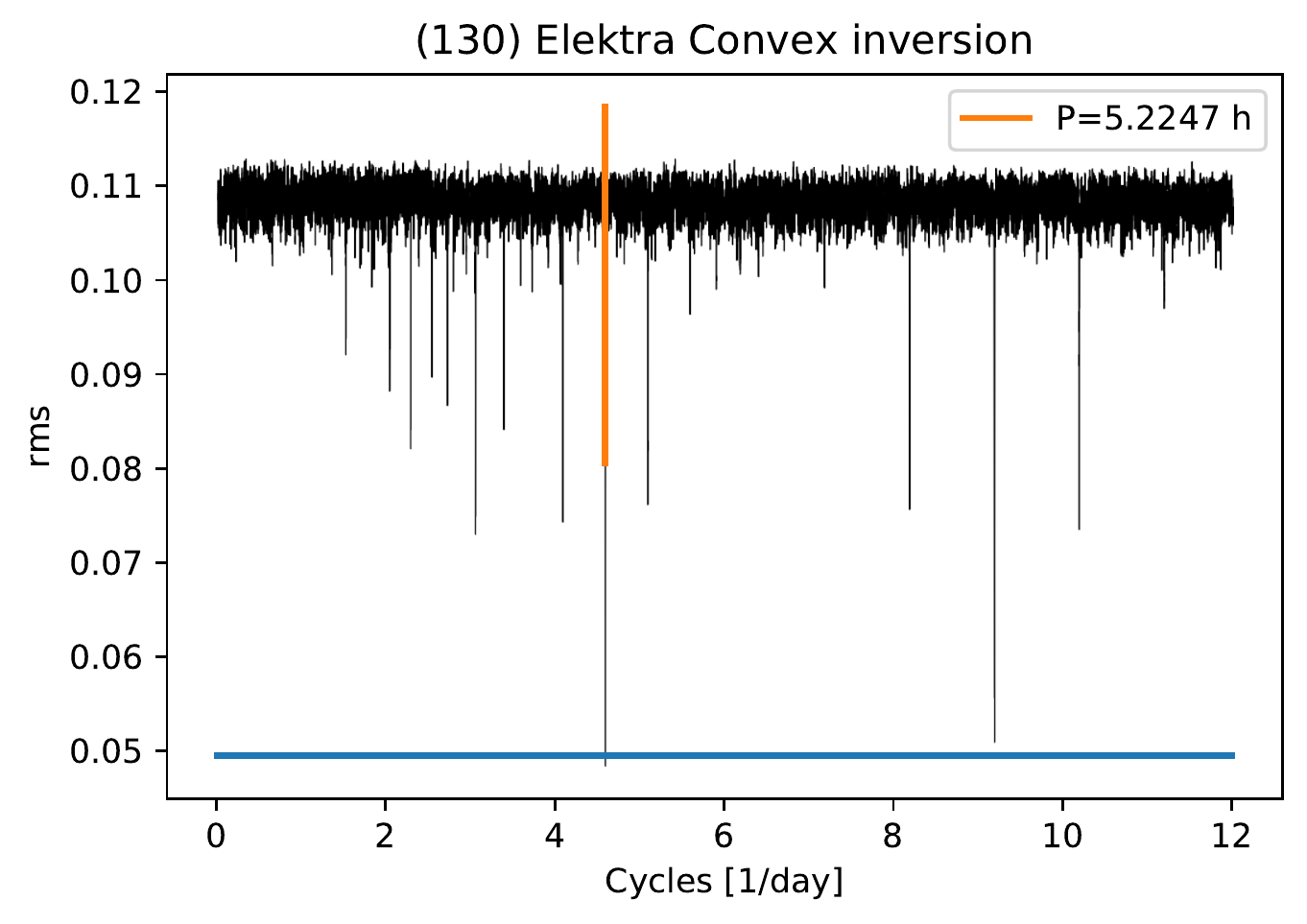}\includegraphics{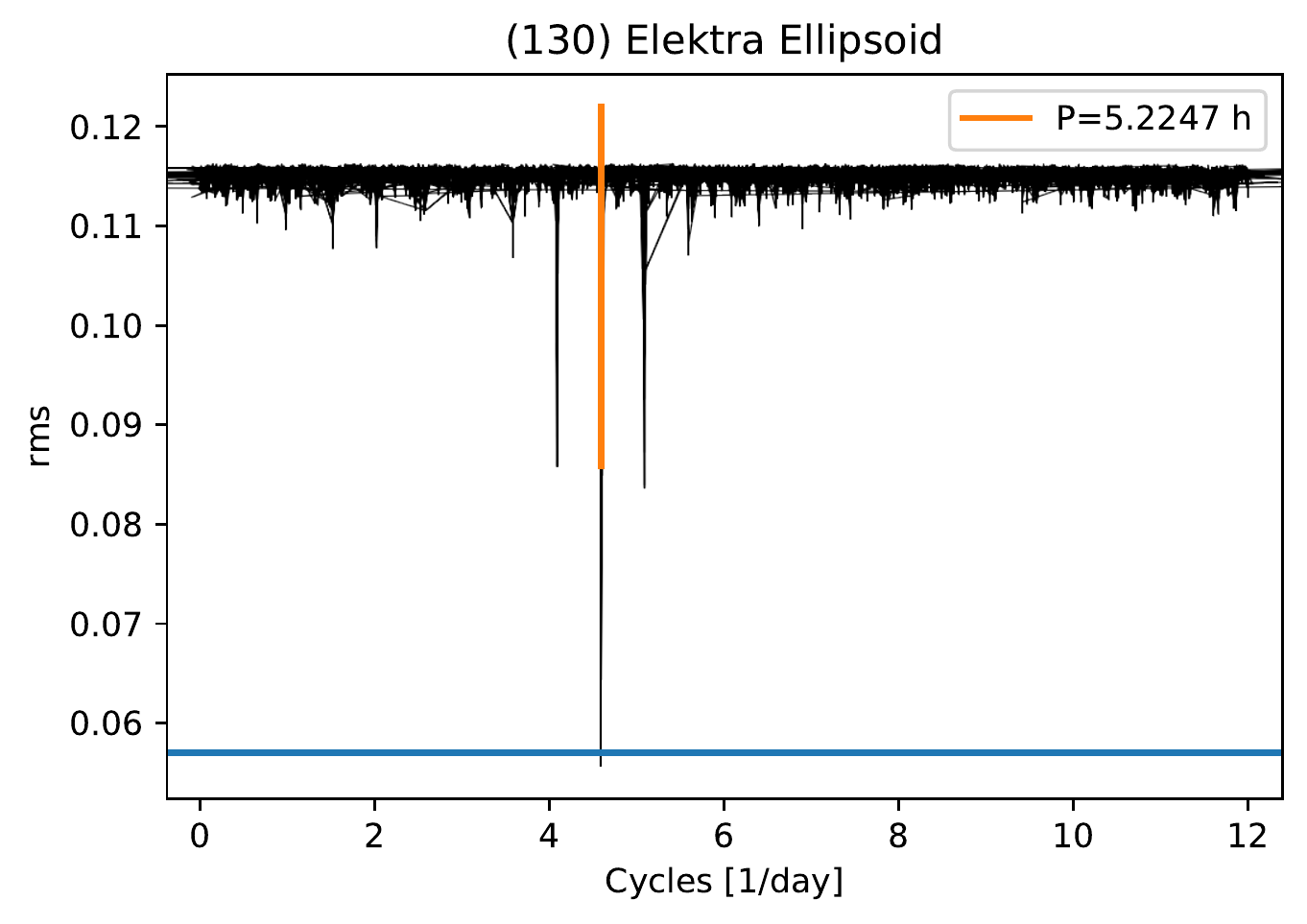}}\\
\end{center}
\caption{\label{fig:ElektraPer}Periodograms (in the frequency domain) of (130) Elektra based on convex inversion (left) and tri-axial ellipsoids (right). Each minimum represents one trial run sampling all the local minima at fixed rotation period. The vertical line indicates the best-fitting value (consistent with the LCDB period of 5.225 h). The horizontal line represents the $\chi^2$ threshold defined by Eq.~\ref{eq:chi2limit}. There are more than 110,000 trial periods for Elektra (see Eq.~\ref{eq:deltaP}). The prominent side minima near $P$ and $P/2$ are the aliases at $\pm$0.5 cycles day$^{-1}$ around $P$ and $\pm$1 cycles day$^{-1}$ around $P/2$. Convex inversion models generically show more periodogram structure than the ellipsoidal models.}
\end{figure*}

We applied the convex inversion (CI) of \citet{Kaasalainen2001a} and \citet{Kaasalainen2001b} to all asteroids with more than 60 individual measurements (5,283 asteroids). This requirement reflects the minimum number of free parameters in the model (55):  49 for shape parameterization, 3 for rotation state (sidereal rotation period, orientation of the spin axis), and 3 phase function parameters (Eq.~\ref{eq:phase}). CI is a gradient-based inversion algorithm that converges to the nearest local minimum given the initial values of the rotation state parameters. We needed to run the minimization multiple times to ensure that we did not miss the global minimum. In practice, the shape modeling consists of four consecutive steps.

In the first step we scan the period interval of 2--1,500 hours. All the asteroids in our sample are large enough to be viewed as rubble piles that cannot rotate faster than the critical rotation period of $\sim$2 hours \citep{Pravec2000}. Rotation periods longer than 1,500 hours are extremely rare and difficult to derive even from sky-survey data. Moreover, these bodies likely rotate in the non-principle axis regime, which we cannot properly model by our method \citep{Pravec2005}. To find the true rotation period, we need to sample a parameter space that is densely filled with local minima. The difference between two local minima in the rotation period parameter space $\Delta P$ can be approximated as
\begin{equation}\label{eq:deltaP}
\frac{\Delta P}{P} = \frac{1}{2}\frac{P}{T},
\end{equation}
where $T$ is the time span of the data \citep{Kaasalainen2001b}. Therefore, we run the shape models for all periods in the 2--1,500 h interval separated by 0.5$\Delta P$. For each trial period, we run the shape models with ten different initial pole orientations, isotropically distributed on a sphere.

In the second step, if we find a single rotation period that fits the data significantly better than all the other period values, we run the convex inversion with this period and a finer grid for the initial pole orientations. Our chi-square threshold is defined by
\begin{equation}\label{eq:chi2limit}
\chi^2_{\mathrm{tr}} = \left(1+p\sqrt{\frac{2}{\nu}}\right)\chi^2_{\mathrm{min}},
\end{equation}
where $\chi^2_{\mathrm{min}}$ is the chi-square of the global minimum, $\nu$ is the number of degrees of freedom (number of observations minus the number of free parameters), and $p$ is a weighting factor. We discuss the selection of $p$ in Sec.~\ref{sec:threshold}.

In the third step, if we find one or two pole solutions that fit the data better than the remaining ones (again by using Eq.~\ref{eq:chi2limit}), we use the convex inversion algorithm to obtain a unique solution with these poles and rotation period as starting values.

In the fourth step the shape model is represented by a set of areas and their normals \citep{Kaasalainen2001a}, and the vertices (the shape is represented as a convex polyhedron) are computed using Minkowski minimization \citep{Kaasalainen1992a, Lamberg1993}.

As in \citet{Durech2018d} and \citet{Durech2020}, we also use a simple tri-axial ellipsoidal model to scan the rotation period parameter space. These models are easily forced to be physical by requiring the semi-major axes to be $a>b>c=1$. The convention is that $c$ is the principal rotation axis that maximizes the moment of inertia. While convex inversion often results in similarly good fits for both the correct period $P$ and its multiples, most commonly $P/2$, light curve inversion with tri-axial ellipsoids usually favors only the correct period $P$ (see Fig.~\ref{fig:ElektraPer}). Fortunately, convex inversion solutions with periods $P/2$ are often nonphysical because they also tend to force the principle rotation axis $c$ to be the longest, which is nonphysical. We only use the ellipsoidal model for the rotation period and then compute the shape using the convex inversion  method. 
\subsection{Period threshold}\label{sec:threshold}

\begin{figure}
\begin{center}
\resizebox{0.99\hsize}{!}{\includegraphics{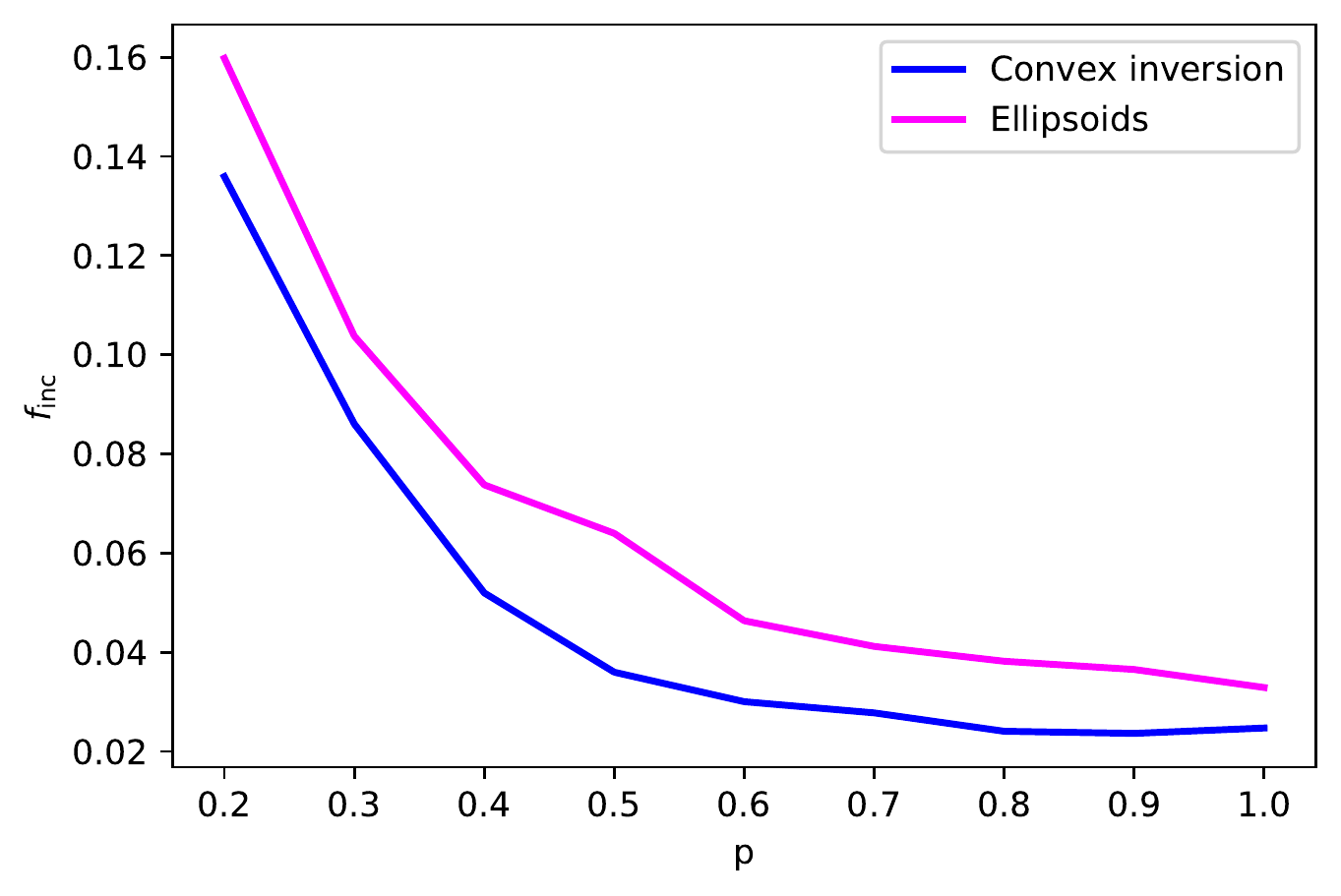}}\\
\end{center}
\caption{\label{fig:falsePer}Dependence of the false solution rate $f_{\mathrm{inc}}$ on the parameter $p$ defined in Eq.~\ref{eq:chi2limit} for both the convex inversion (blue) and tri-axial ellipsoid models (magenta).}
\end{figure}

\begin{figure*}
\begin{center}
\resizebox{1.\hsize}{!}{\includegraphics{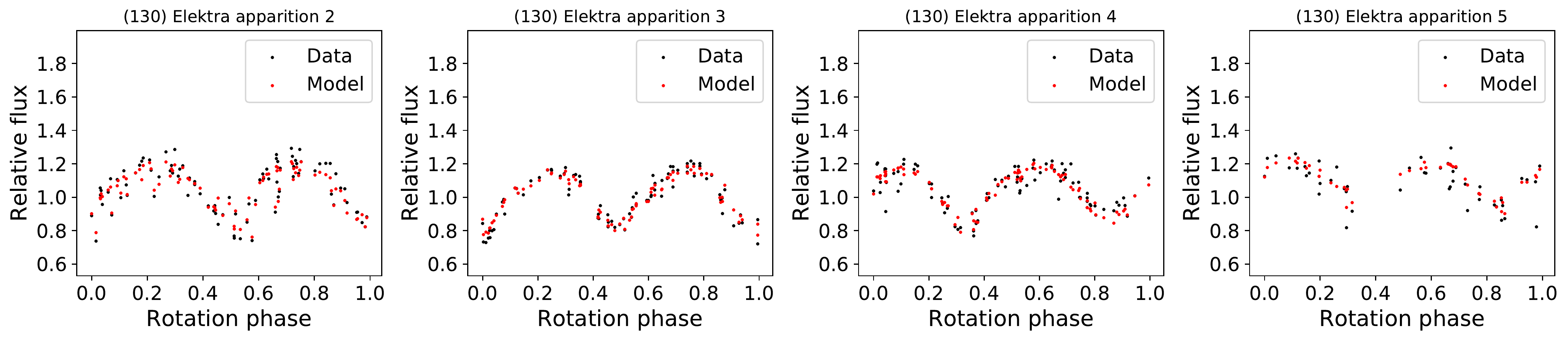}}\\
\end{center}
\caption{\label{fig:130folded} Fit to the ASAS-SN data for asteroid (130)~Elektra. The data is split into four apparitions and is shown  as a function of rotation phase given the sidereal rotation period of 5.22466~h. The model for each apparition shows scatter because the geometry is also changing during each apparition. The phase reference time is defined independently for each apparition.}
\end{figure*}

Our criterion for the uniqueness of the best-fitting rotation period is defined by Eq.~\ref{eq:chi2limit}, where we need to select the optimal value of the $p$ factor. We do this using comparisons to the asteroid Lightcurve Database \citep[LCDB,][]{Warner2009} compilation of asteroid physical properties, including rotation periods. Each period is ranked by a flag that indicates its reliability. For our analysis we selected only the most reliable values with flags ``3'' and ``3$-$''. Figure~\ref{fig:falsePer} shows the fraction $f_\text{inc}$ of the ASAS-SN periods that disagree with the previously reported values as a function of $p$ for the two fitting methods. We omit cases where the LCDB period is twice the ASAS-SN period. For $p<0.5$, $f_{\mathrm{inc}}$ increases rather rapidly. So we select $p=0.5$ as an optimal value in the CI and $p=0.6$ in the tri-axial ellipsoid shape analyses. 

For these choices, about 3\% of periods derived by CI from ASAS-SN data are inconsistent with the LCDB periods. We discuss several individual cases in Sect.~\ref{sec:individual}. It is important to keep in mind that not all flag 3 and 3- periods in LCDB will be correct. \citet{Durech2016} and \citet{Durech2020} found multiple examples of incorrect estimates and $f_{\mathrm{inc}}$ on the order of a few percent is not particularly concerning. The incorrect periods were generally for asteroids with $P>20$ h, where determinations are more challenging for classical short duration, dense sampling light curve observations. The $f_{\mathrm{inc}}$ ratio for the tri-axial ellipsoid approach is higher by about 2\% than for the convex inversion approach. The minor increase is likely due to the ellipsoidal approach being less physically realistic.

We also   see cases where the ASAS-SN period is clearly incorrect. For example, a few tens of the brightest asteroids have at least partly saturated photometry, which sometimes result  in best-fitting period values of 24 or 48 h. We rejected these objects from our analysis. Additionally, we also identified cases with a relatively low number of detections ($<100$) and ``suspicious'' periodograms. Usually, good periodograms are qualitatively similar to the one of asteroid (130)~Elektra (Fig.~\ref{fig:ElektraPer}); most of these solutions cluster along the line of constant maximum rms and only several deep and sharp minima are present, usually in simple ratios. Suspicious periodograms have larger scatter and a more random distribution of minima. We also removed these solutions from our analysis. Finally, many solutions with inconsistent LCDB periods were rejected by our additional reliability tests.

We   only accepted a solution if it had at most two pole solutions passing the statistical threshold of Eq.~\ref{eq:chi2limit}, and if the two poles were the so called mirror solutions \citep[similar ecliptic latitude and an ecliptic longitude difference of $\sim180$ degrees,][]{Kaasalainen2006}. The shape solution  also needed to be physically plausible, with the body rotating along its shortest axis. Therefore, we computed the axis with the maximum momentum of inertia of the body \citep{Dobrovolskis1996} and compared it with the position of the spin axis. 

We evaluated the models using the dependence of the brightness on phase angle (Fig.~\ref{fig:130data}) and rotation phase (Fig.~\ref{fig:130folded}), which we illustrate using asteroid (130) Elektra. In both cases the mean brightness is normalized to unity. Because the geometry changes with apparition, the phase-folded light curves change with each apparition. The models for each apparition also show scatter due to changes in geometry during the individual apparitions. Similarly, the scatter in the data is a combination of  the photometric errors and the changes in geometry. Moreover, the folded plots also illustrate the sufficient coverage of the rotation phase by the data and that the residuals are significantly smaller than the lightcurve amplitude.

\subsection{New shape solutions from ASAS-SN data}\label{sec:individual}

\begin{figure*}
\begin{center}
\resizebox{0.8\hsize}{!}{\includegraphics{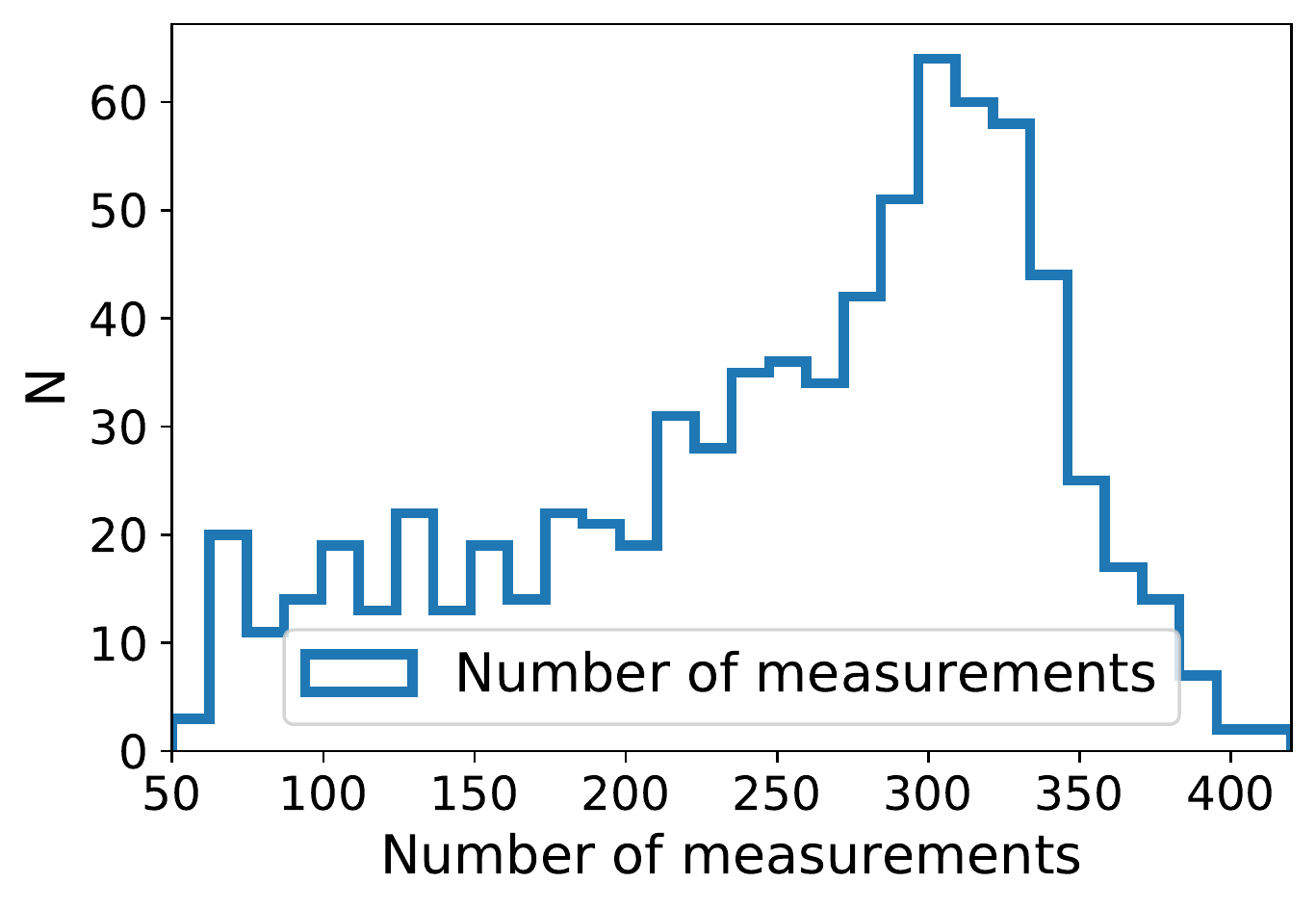}\includegraphics{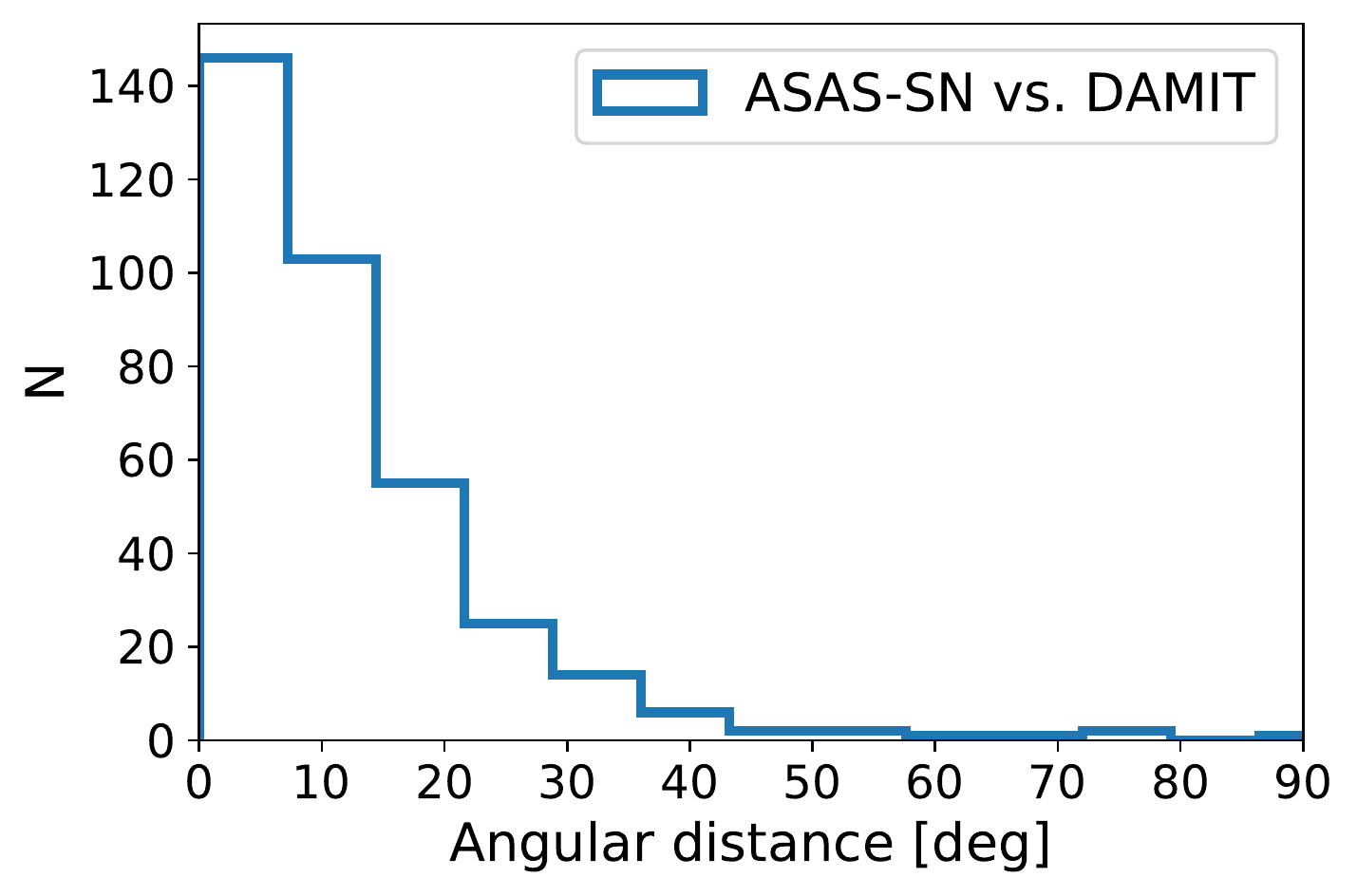}}\\
\end{center}
\caption{\label{fig:statistics}Data amount and pole consistency for ASAS-SN solutions. \textit{Left}: Histogram of the number of ASAS-SN $V$-band measurements for the asteroids with successful shape solutions. \textit{Right}: Angular distance between the ASAS-SN and DAMIT pole solutions.}
\end{figure*}

\begin{figure*}
\begin{center}
\resizebox{0.8\hsize}{!}{\includegraphics{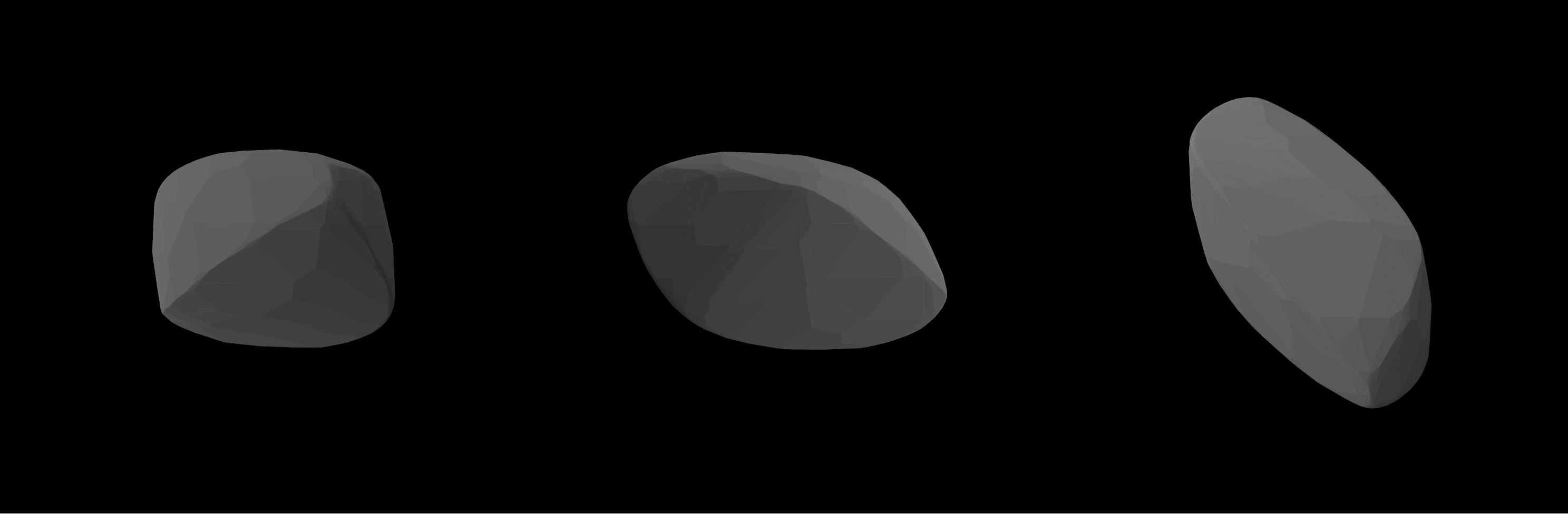}}\\
\end{center}
\caption{\label{fig:shape}First convex shape model of asteroid (1532)~Inari. The left and center views are equatorial with a 90$^\circ$ rotation, while the right view is pole-on.}
\end{figure*}

From the  5,283 asteroids with more than 60 photometric measurements, we derived unique spin state and shape solutions for 760 asteroids. This corresponds to a success rate of about 15\%, which is significantly higher than the success rate from the previous wide-area surveys such as ATLAS \citep[$\sim$3\%,][]{Durech2020}, Gaia DR2 \citep[$\sim$3\%,][]{Durech2018d}, and Lowell alone or in combination with WISE \citep[$\lesssim$1\%,][]{Durech2016, Durech2018c}.

We obtained new shape model determinations for 163 individual asteroids. A typical example is shown in Fig.~\ref{fig:shape} (asteroid 1532 Inari). We list their physical properties in Table~\ref{tab:models_new}. The solutions for asteroids with shape models already included in DAMIT are provided in Table~\ref{tab:models_damit}. Figure~\ref{fig:statistics} compares our pole directions and those reported in DAMIT. The difference is rarely larger than 30$^\circ$, which suggests a good consistency between our solutions and previous estimates. Additionally, we derived sidereal rotation periods for 69 asteroids for the first time (Table~\ref{tab:periods_new}). However, pole orientations for these asteroids remained ambiguous. Clearly, more photometric data are still needed here. Interestingly, the fraction of slowly rotating asteroids is large for these 69 asteroids. 

We identified four solutions that were inconsistent with those from DAMIT. Usually, the previous shape modeling was performed over a short interval spanning the rotation period reported in the LCDB database. However, in some cases, either this rotation period was   shown to be incorrect or we now prefer a different one (asteroids 1040 and 2962). In two cases, we derived a rotation period that is twice that of DAMIT (asteroids 1461 and 16009).

Within our new shape and spin state solutions, 23 periods are inconsistent with the LCDB periods by more than 5\%. Only one has the highest  LCDB reliability flag of 3 (asteroid 974). With only one exception (asteroid 1447~Utra), the new periods are larger and typically $>$100\,h. It is not surprising that these cases generally have lower reliability flags and long periods.

\begin{figure*}
\begin{center}
\resizebox{0.99\hsize}{!}{\includegraphics{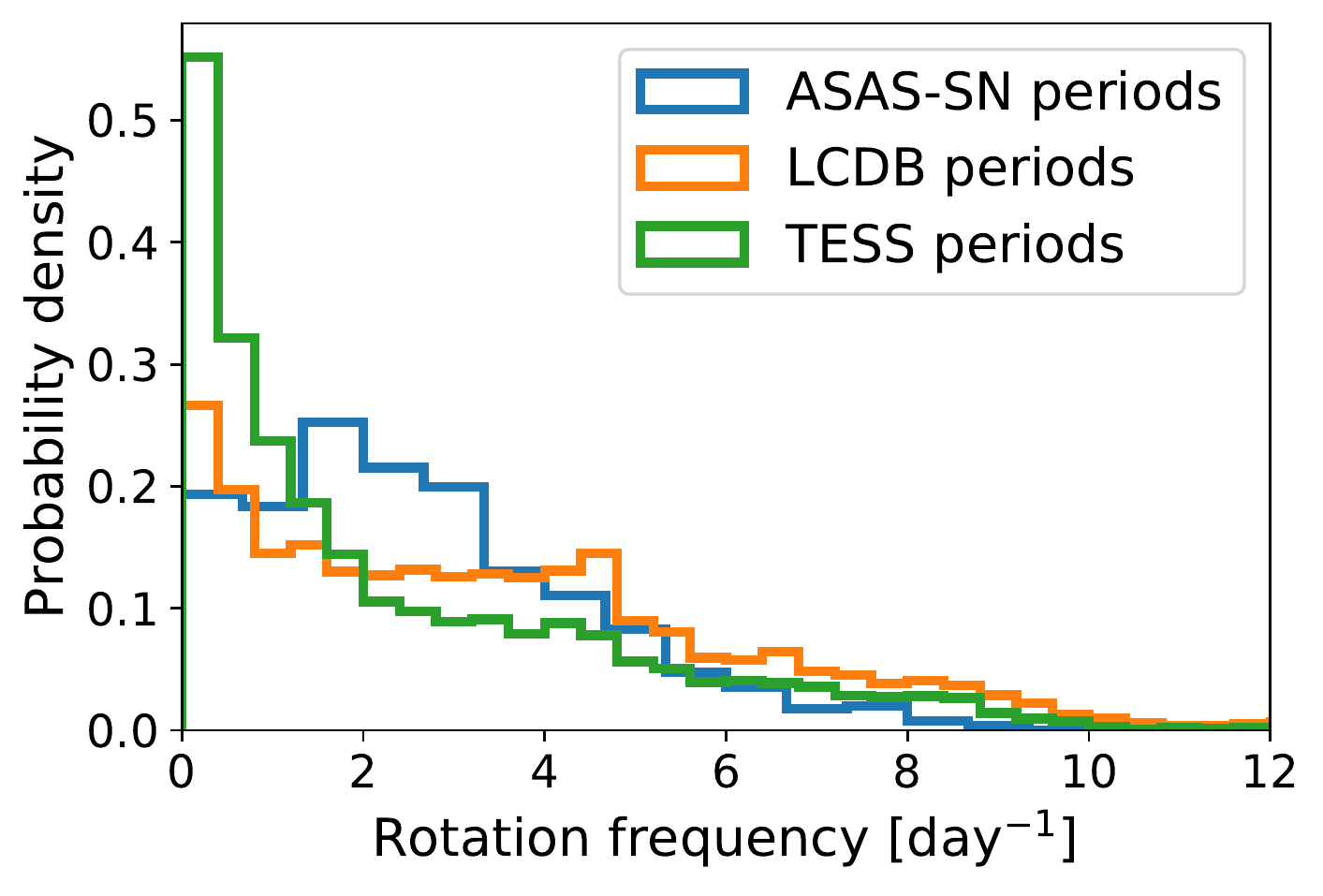}\includegraphics{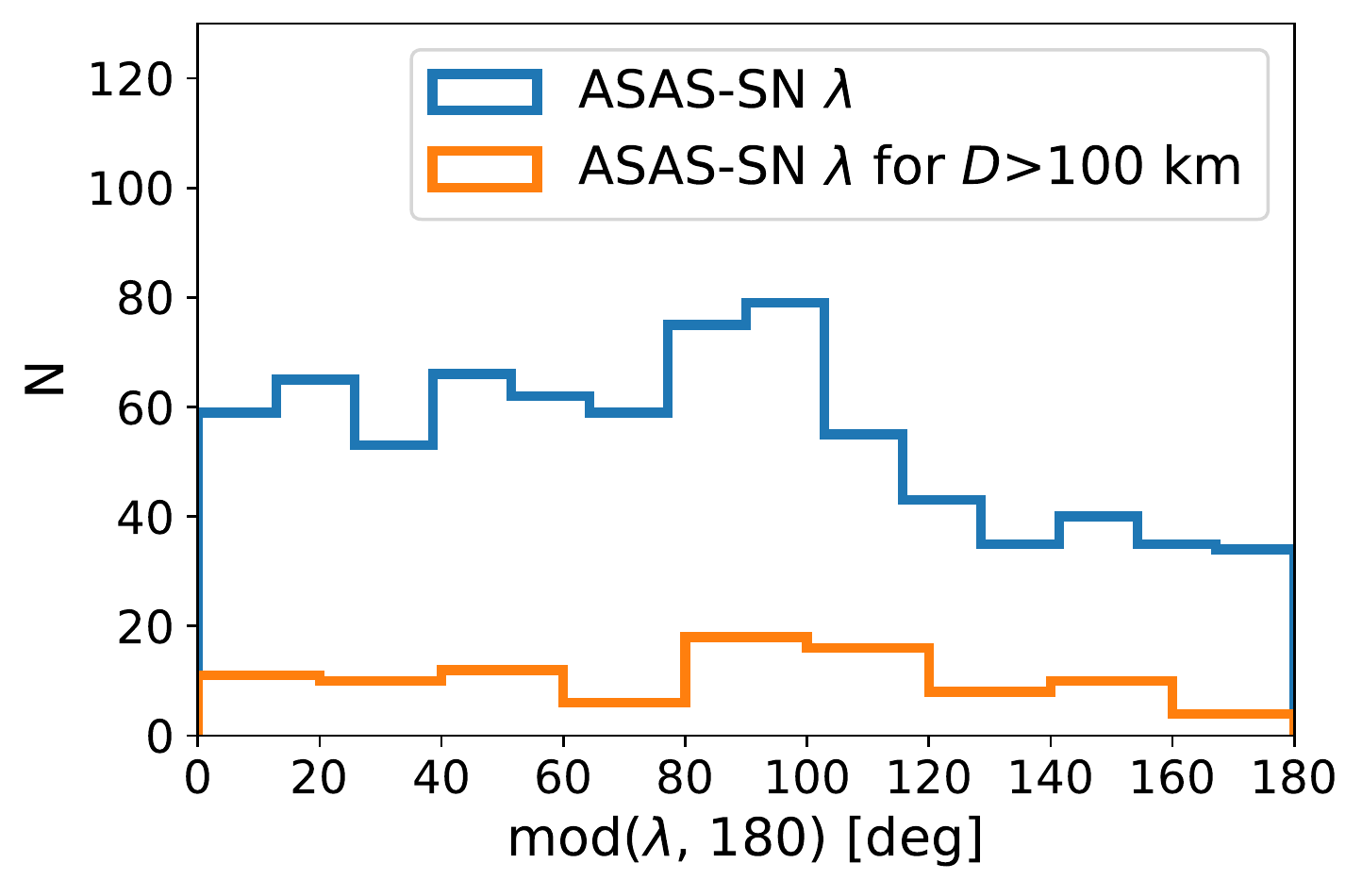}\includegraphics{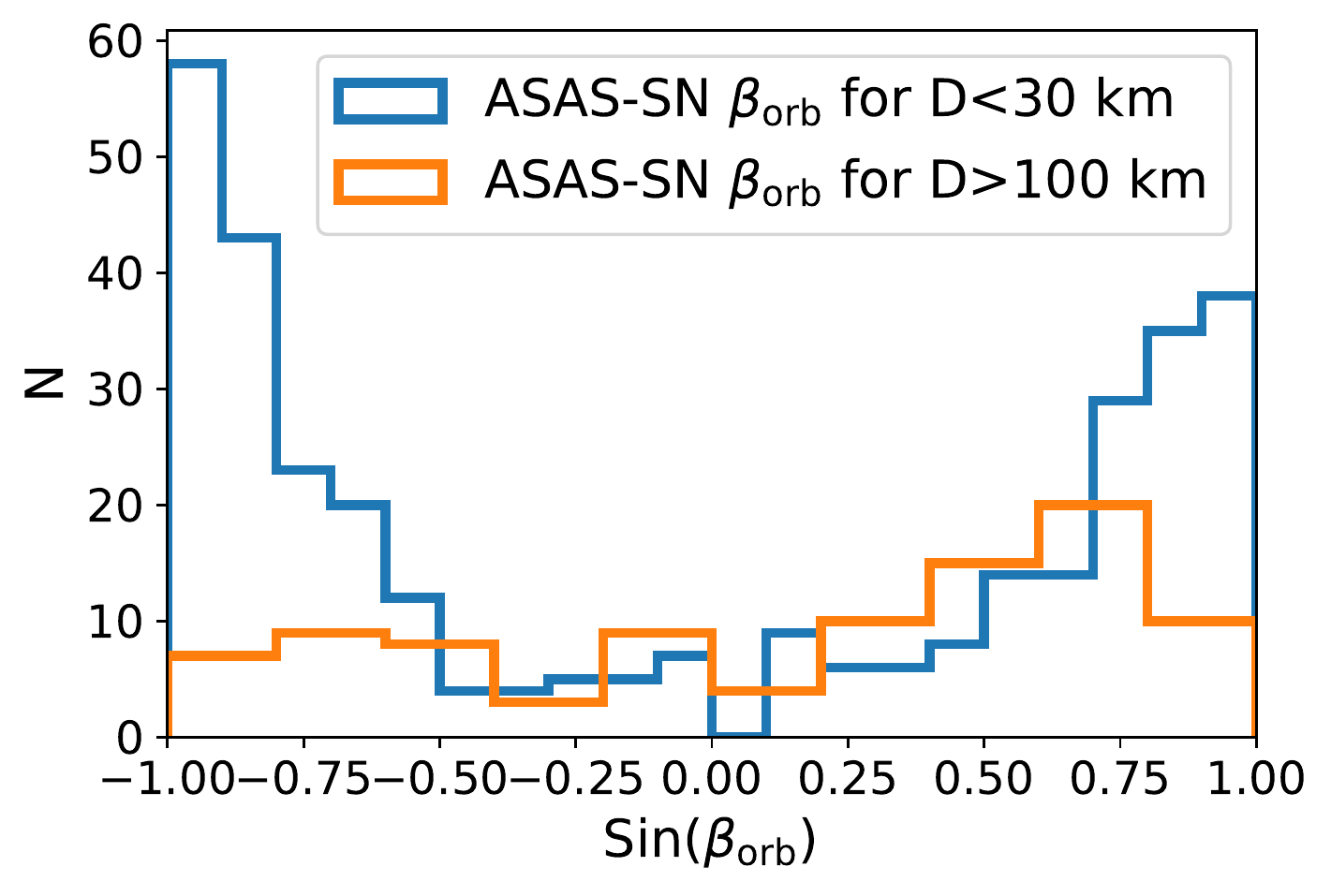}}\\
\end{center}
\caption{\label{fig:spinstate}Rotation period and pole direction distributions. \textit{Left}: Normalized period distributions for the ASAS-SN models (760), \textit{TESS} periods \citep[9,912,][]{Pal2020}, and  LCDB periods \citep[31,280,][]{Warner2009}. \textit{Middle}: Distribution of pole ecliptic longitudes $\lambda$ for all asteroids and asteroids $>100$\,km derived from the ASAS-SN data. \textit{Right}: Distribution of orbital pole latitudes $\beta_\mathrm{orb}$ derived from ASAS-SN data for asteroids smaller than 30 km and for asteroids larger than 100 km.}
\end{figure*}

\begin{figure*}
\begin{center}
\resizebox{0.8\hsize}{!}{\includegraphics{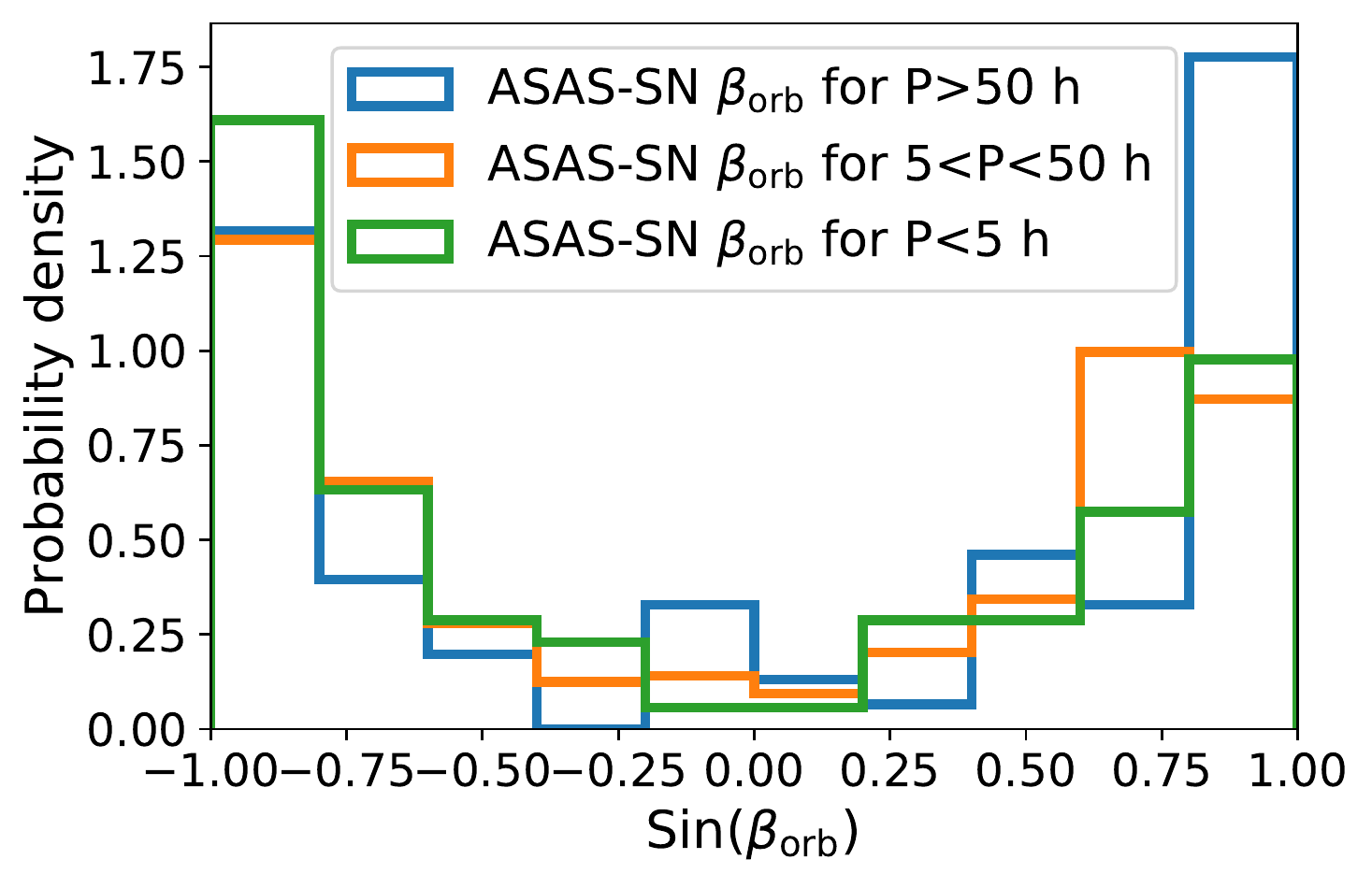}\includegraphics{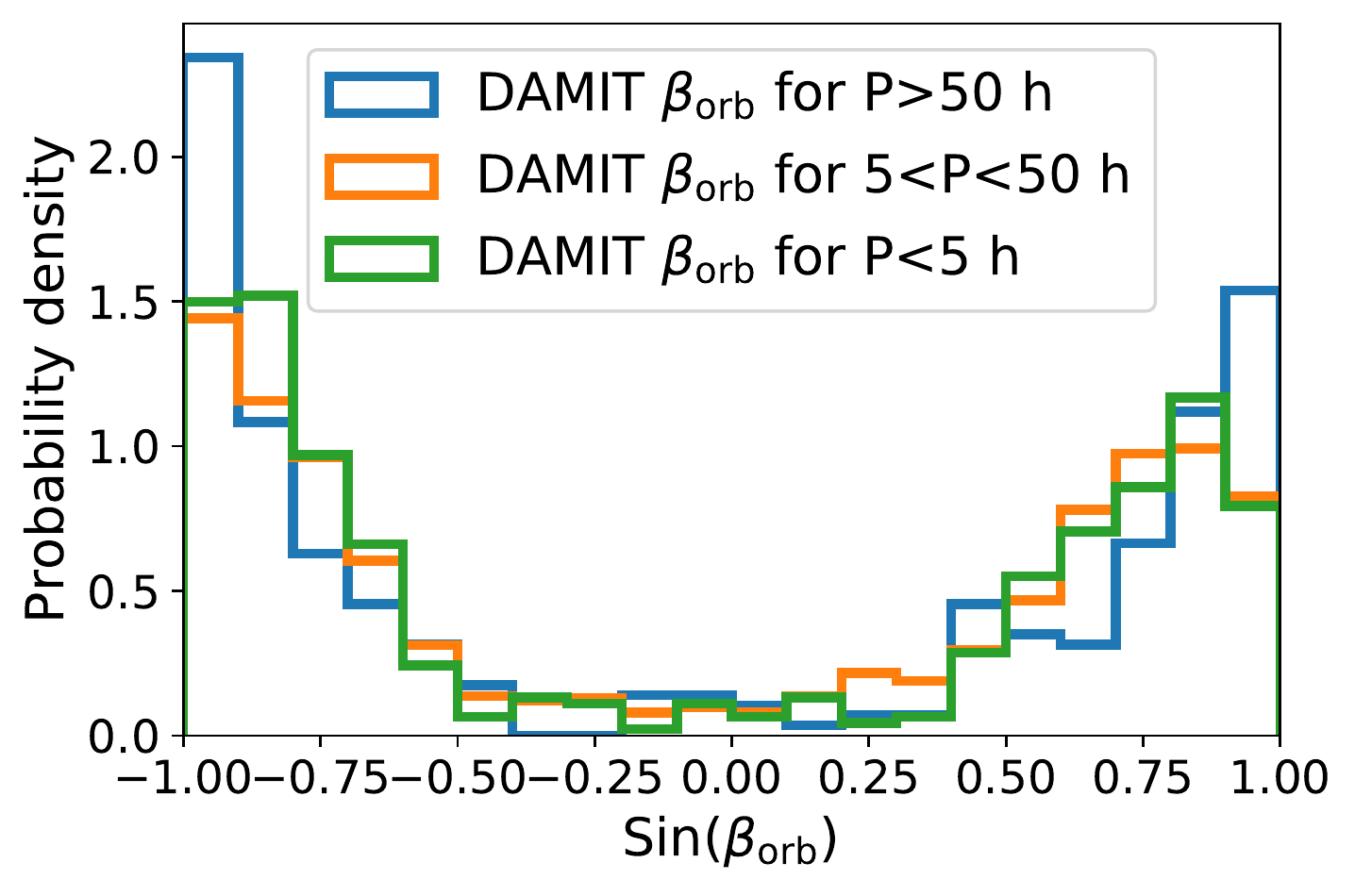}}\\
\end{center}
\caption{\label{fig:spinstateSlow}Distribution of orbital pole latitudes $\beta_\mathrm{orb}$ of asteroids with $D<50$~km for different rotation period ranges derived from ASAS-SN data (left) and from the DAMIT database (right).}
\end{figure*}

In Figure~\ref{fig:spinstate} we show the distribution of rotation periods obtained from ASAS-SN  compared to all those available from LCDB.  ASAS-SN data are very efficient in identifying long-period rotators due to the long time span and high cadence of the survey. We discuss this issue in Sect.~\ref{sec:slowExcess}. Figure~\ref{fig:spinstate} also shows the distribution of pole ecliptic longitudes $\lambda$ and orbital latitudes $\beta_\text{orb}$ for different asteroid sizes, revealing the signatures of YORP in small asteroids. We give more details about this finding in Sect.~\ref{sec:spinanis}. Finally, in Figure~\ref{fig:spinstateSlow}, we show the distribution of $\beta_\text{orb}$ as a function of rotational period. We discuss the implications of our findings for slow rotators in Sect.~\ref{sec:spinslow}.

\section{Discussion}\label{sec:discussion}

ASAS-SN data represent the first sky survey with a fraction of successful shape determinations (15\%) exceeding ten percent. This is an order of magnitude improvement over previous efforts based on sparse-in-time datasets. This is partly because of the better photometric accuracy, but also because we limited our analysis to epochs with photometric errors $<0.15$\,mag. The other analyses typically examined a large number of asteroids close to the detection limit. Such data have photometric uncertainties exceeding 0.15 mag. Moreover, information about the uncertainties is often missing, which precludes filtering based on measurement uncertainty. Shape modeling performed on noisy data usually does not provide a unique solution, which decreases the success rate. 

We also tested shape modeling with the photometric uncertainty threshold reduced to $0.1$\,mag. As expected, the resulting dataset contains  fewer asteroids and fewer measurements. However, the brighter asteroids are almost unaffected as their measurements usually have accuracy better than 0.1 mag. The faint asteroids are more affected;  the photometric accuracy is often in the 0.1--0.15\,mag range. Considering the shape modeling results, the two datasets are rather comparable. In general, adding noisier photometry (i.e., using a threshold of 0.15\,mag instead of 0.1\,mag) resulted in a small number of missed solutions due to the poorer photometric accuracy, but also produced some additional solutions due to the larger number of measurements available for fainter asteroids. We prefer deriving new solutions for fainter asteroids (i.e., threshold of 0.15 mag) rather than confirming more solutions that are already known.

\subsection{Excess of slow rotators}\label{sec:slowExcess}

\citet{Pal2020} derived 9,912 rotation periods using \textit{TESS} data and found a clear excess of slow ($P\gtrsim12$ h) rotators compared to existing compilations (their Fig. 7), which suggests these objects were  underrepresented in previous ground-based surveys. Similar excesses were also found by \citet{Kiss2019} and \citet{Durech2020}. This is due to the difficulty of determining long periods using the traditional ground-based observations on which most of the LCDB periods relied \citep{Marciniak2015,Marciniak2018}. 

Our sample of 920 ASAS-SN periods show an excess of asteroids with periods between $\sim$7 and 24~h (see Fig.~\ref{fig:spinstate}) when compared to the \textit{TESS} \citep{Pal2020} and the LCDB period distributions. We used the most recent version of the LCDB database with 31,280 period estimates, including all the 9,912 \textit{TESS} periods from \citet{Pal2020}. Clearly, periods $>$24~h are not as frequent as in the pure \textit{TESS} sample. This likely originates in the smaller success rate of shape model determinations for asteroids with these longer periods because the geometry coverage is probably not sufficient. ASAS-SN data seem to be quite useful for deriving shape models and spin states for asteroids with rotation periods of 7--25~h.

We recall  that all the currently available rotation period samples are affected by complicated (often unknown or poorly understood) biases. Therefore, these biases should require proper consideration before any more conclusive interpretations are drawn.

\subsection{Spin axis anisotropy}\label{sec:spinanis}

The clustering of spin vectors for asteroids smaller than $\sim$30 km was already illustrated and explained by \citet{Hanus2011}, and further confirmed in larger datasets \citep{Hanus2013a,Durech2016}. The thermal forces known as the YORP effect \citep[][]{Bottke2006, Vokrouhlicky2003} drive a gradual evolution of the spin axis to be perpendicular to the orbital plane. The pole latitudes of the spin axis should cluster near $|\sin \beta_\text{orb}| =1$ in our notation. We note that CI provides the pole orientation in the ecliptic coordinate frame ($\lambda$, $\beta$) and we need to transform it into the orbital frame of the asteroid, where the obliquity of the pole is $\epsilon = 90-\beta_\mathrm{orb}$. As the orbital inclinations of typical main-belt asteroids are small ($\lesssim$20$^\circ$), the differences between $\beta$ and $\beta_\mathrm{orb}$ are rather small as well.

We clearly see the latitude clustering in our data for asteroids smaller than 30~km (Fig.~\ref{fig:spinstate}), where the YORP timescale is shorter than the reorientation timescale due to collisions \citep{Hanus2011}. Large asteroids ($D>100$~km) are not significantly affected by the YORP effect and their pole latitudes are randomly distributed (Fig.~\ref{fig:spinstate}), as  expected for a collisionally dominated population. However, there is a statistically significant excess of prograde rotators ($\beta_\mathrm{orb}>$0$^\circ$) that can be explained by accretion of the pebbles into protoplanets in the prograde-rotating gaseous disk \citep{Johansen2010}.

\citet{Bowell2014a} noted that their distribution of ecliptic longitudes was anisotropic. We see a similar distribution of ecliptic longitudes (Fig.~\ref{fig:spinstate}) with an excess of asteroids at longitudes of 60--110$^\circ$. The anisotropy also seems to be present for large asteroids ($>100$~km, Fig.~\ref{fig:spinstate}), which suggests it is independent of the size. Interestingly, our latitude distribution for asteroids larger than 100~km is based on 96 solutions, almost  half of the whole population (about 200 bodies). The anisotropic distribution of the ecliptic longitudes remains unexplained.

The observed longitude and latitude distributions are affected by observing biases. \citet{Hanus2011} performed debiassing of these distributions and showed that the debiased latitude distribution for small asteroids is still clearly anisotropic. Moreover, the observation bias in the longitude is negligible, and thus cannot explain the observed anisotropy.

\subsection{Spin states of slow rotators}\label{sec:spinslow}

Asteroids having $P>50$~h make up a peculiar group, and are not yet fully understood. There are only about $300$ such systems in DAMIT and here we have added an additional $45$. The general consensus is that these bodies were de-accelerated by the YORP thermal effect \citep[e.g.,][]{Capek2004, Pravec2008}. This is in  agreement with the population being composed only of small asteroids ($D<50$~km). Larger asteroids have a Maxwellian distribution of rotation periods consistent with their collisional origin \citep{Pravec2008}. However, alternative theories have also been put forward. For instance, some fragments of disruptive collisions could have slow spin rates or the slow rotation could originate in a synchronous binary that split apart. Moreover, a large fraction of slow rotators are in an excited rotation state, the so-called tumbling mode \citep{Harris1980, Pravec2005, Pravec2008}. There are two main scenarios proposed for the excitation: YORP spin-up eventually self-triggering the tumbling \citep{Vokrouhlicky2007} or YORP spin-up followed by  a noncatastrophic collision \citep{Henych2013}. 

In Fig.~\ref{fig:spinstateSlow} we show the pole latitude $\beta_\mathrm{orb}$ distributions for three different bins in rotation period, including slow rotators ($P>50$~h), for  the ASAS-SN solutions and for those in DAMIT. For both datasets the slow rotators have a larger number of poles near $\beta_\mathrm{orb}=\pm$90$^\circ$ compared to the rest of the population, so there are more spin vectors in the YORP end states. The natural explanation would be that the population of slow rotators is more strongly affected by YORP. It takes more time to slow the period to $P>50$\,h, and so there is enough time to reach the YORP spin vector end state. A minor complication to this scenario is the presence of  secular spin-orbit resonances in which the spin vectors could be temporarily trapped, as in the case of the Koronis family \citep{Slivan2002, Vokrouhlicky2003}. These resonances are likely responsible for the asymmetry in the $\beta_\mathrm{orb}$ distribution for asteroids with $D<30$~km because they affect only the prograde-rotating asteroids. However, as the excess of spin vectors in the end states of slow rotators is present for both prograde and retrograde bodies, the resonances cannot be the main cause. Another process that acts against the YORP evolution is noncatastrophic collisions. As the slow rotators have a rather small moment of inertia compared to the faster rotators, even a small collision can significantly affect the rotation state. For instance, an impact can change the orientation of the spin axis, the value of rotation period (in both directions) or even excite the rotation. More spin state data combined with dynamical studies should shed more light into the processes that are the most active in shaping the population of slow rotators.

\section{Conclusion}\label{sec:conclusions}

We showed that ASAS-SN $V$ band photometry is a good tool for the physical characterization of asteroids by deriving shapes and rotation properties for 760 asteroids, including 163 new determinations. We validated our results by comparison with determinations based on independent photometric datasets. We also revised incorrect period estimates for several asteroids.

We obtained a statistical sample of asteroid physical properties that was sufficient to confirm and expand on several previously known trends: 
(i)~the underrepresentation of slow rotators in current databases of rotation period estimates (i.e., LCDB) due to observation biases,
(ii)~the anisotropic distribution of ecliptic pole-latitudes due to nongravitational forces (YORP), and
(iii)~an anisotropic distribution of ecliptic pole-longitudes of unknown origin.

Most of the successful solutions are for asteroids with more than 200 measurements (Fig.~\ref{fig:statistics}), which means that the data cover at least 1,500 days (Fig.~\ref{fig:asasStat}). Clearly, a good coverage of observing geometries is essential for a successful shape model determination. The asteroid also has to be bright enough during the apparition when it has the faintest apparent $V$-band magnitude (the least convenient geometry, when the asteroid is at aphelion). Asteroids that are too faint at this apparition are likely   too faint or have very noisy photometry at the other apparitions. 

The newer ASAS-SN $g$-band data will be quite useful for the shape modeling as they are about one magnitude deeper. Apparitions with poor photometric accuracy or brightness at or slightly below our $16.5$ mag threshold in this paper will have useful $g$-band measurements, which could increase the number of successful solutions by a factor of at least 2, particularly since the number of asteroids  rises from $\sim$20,000 that reach $V \le 16.5$\,mag to $\sim$55,000 that reach $V \le  17.5$\,mag. Another benefit of the $g$-band data is that one-day aliasing will likely be minimized because ASAS-SN now has units at  four different longitudes (at least two for everywhere on the sky). Finally, the full potential of ASAS-SN data lies in the possibility of combination with other datasets, as has been done with others surveys in DAMIT.

\begin{acknowledgements}
We thank the Las Cumbres Observatory and its staff for its continuing support of the ASAS-SN project. ASAS-SN is supported by the Gordon and Betty Moore Foundation through grant GBMF5490 to the Ohio State University, and NSF grants AST-1515927 and AST-1908570. Development of ASAS-SN has been supported by NSF grant AST-0908816, the Mt. Cuba Astronomical Foundation, the Center for Cosmology and AstroParticle Physics at the Ohio State University, the Chinese Academy of Sciences South America Center for Astronomy (CAS-SACA), the Villum Foundation, and George Skestos. 

The work of JH and OP has been supported by INTER-EXCELLENCE grant LTAUSA18093 from the Ministry of Education, Youth, and Sports. The research of JH has been additionally supported by the Czech Science Foundation through grant 20-08218S and by Charles University Research program No. UNCE/SCI/023. The research of OP has been supported also by Horizon 2020 ERC Starting Grant `Cat-In-hAT' (grant agreement no. 803158). 
BJS, CSK, and KZS are supported by NSF grant AST-1907570. BJS is also supported by NASA grant 80NSSC19K1717 and NSF grants AST-1920392 and AST-1911074. CSK and KZS are supported by NSF grant AST-181440.

Support for TW-SH was provided by NASA through the NASA Hubble Fellowship grant HST-HF2-51458.001-A awarded by the Space Telescope Science Institute, which is operated by the Association of Universities for Research in Astronomy, Inc., for NASA, under contract NAS5-26555.

This research has made use of the Minor Planet Physical Properties Catalogue (MP3C) of the Observatoire de la Côte d’Azur and the IMCCE's Miriade VO tool.

\end{acknowledgements}

\bibliography{mybib}
\bibliographystyle{aa}

\newpage
\onecolumn
\begin{appendix}

\section{Additional tables}

\begin{table*}[h]
\caption{\label{tab:data}Example of data stored at the CDS (Strasbourg astronomical Data Center). The table provides each asteroid's number and name, the Julian date, the $V$ band magnitude and uncertainty, and a flag indicating whether the datapoint was used in the light curve inversion (1 for yes, 0 for no).}
\centering
\begin{tabular}{rrrr rr}
\hline 
\multicolumn{1}{c} {Number} & \multicolumn{1}{c} {Name} & \multicolumn{1}{c} {JD} & \multicolumn{1}{c} {$V$ mag} & \multicolumn{1}{c} {$V$ mag unc} & \multicolumn{1}{c} {CI flag} \\
\hline\hline
1 & Ceres & 2456598.148220 & 9.009 & 0.002 & 1 \\
1 & Ceres & 2456601.147060 & 9.151 & 0.002 & 1 \\
1 & Ceres & 2456619.125000 & 9.306 & 0.002 & 1 \\
1 & Ceres & 2456622.110740 & 9.189 & 0.002 & 1 \\
1 & Ceres & 2456625.116330 & 8.933 & 0.002 & 1 \\
1 & Ceres & 2456626.113730 & 9.029 & 0.002 & 1 \\
1 & Ceres & 2456642.077830 & 9.117 & 0.002 & 1 \\
1 & Ceres & 2456657.056800 & 8.865 & 0.002 & 1 \\
1 & Ceres & 2456659.085710 & 8.914 & 0.002 & 1 \\
1 & Ceres & 2456675.054510 & 8.859 & 0.002 & 1 \\
1000 & Piazzia & 2455937.957240 & 17.015 & 0.205 & 0 \\
1000 & Piazzia & 2455999.891650 & 16.677 & 0.105 & 1 \\
1000 & Piazzia & 2456003.909970 & 17.154 & 0.131 & 1 \\
1000 & Piazzia & 2456319.039560 & 15.460 & 0.059 & 1 \\
1000 & Piazzia & 2456326.080760 & 15.373 & 0.044 & 1 \\
1000 & Piazzia & 2456336.952390 & 15.010 & 0.029 & 1 \\
1000 & Piazzia & 2456344.986530 & 15.137 & 0.041 & 1 \\
1000 & Piazzia & 2456354.835490 & 14.700 & 0.026 & 1 \\
1000 & Piazzia & 2456359.880710 & 14.484 & 0.024 & 1 \\
1000 & Piazzia & 2456360.887700 & 14.424 & 0.024 & 1 \\
1000 & Piazzia & 2456371.931610 & 14.352 & 0.023 & 1 \\
1000 & Piazzia & 2456991.695440 & 16.886 & 0.173 & 0 \\
1000 & Piazzia & 2457013.694630 & 16.538 & 0.201 & 0 \\
1000 & Piazzia & 2457183.116350 & 16.350 & 0.190 & 0 \\
\hline
\end{tabular}
\end{table*}

\onecolumn

\begin{longtable}{r@{\,\,\,}l rrrr D{.}{.}{6} rrrrr}
\caption{\label{tab:models_new}Physical properties of asteroids with new shape model determinations based on the $V$ band ASAS-SN data.}\\
\hline 
\multicolumn{2}{c} {Asteroid} & \multicolumn{1}{c} {$\lambda_1$} & \multicolumn{1}{c} {$\beta_1$} & \multicolumn{1}{c} {$\lambda_2$} & \multicolumn{1}{c} {$\beta_2$} & \multicolumn{1}{c} {$P$} & M & $N_{\mathrm{A}}$ & \multicolumn{2}{c} {$P_{\mathrm{LCDB}}$} & \multicolumn{1}{c} {$D$}\\
\multicolumn{2}{l} { } & [deg] & [deg] & [deg] & [deg] & \multicolumn{1}{c} {[hours]} &  &  & \multicolumn{2}{c} {[hours]} & [km] \\
\hline\hline

\endfirsthead
\caption{continued.}\\

\hline
\multicolumn{2}{c} {Asteroid} & \multicolumn{1}{c} {$\lambda_1$} & \multicolumn{1}{c} {$\beta_1$} & \multicolumn{1}{c} {$\lambda_2$} & \multicolumn{1}{c} {$\beta_2$} & \multicolumn{1}{c} {$P$} & M & $N_{\mathrm{A}}$ & \multicolumn{2}{c} {$P_{\mathrm{LCDB}}$} & \multicolumn{1}{c} {$D$}\\
\multicolumn{2}{l} { } & [deg] & [deg] & [deg] & [deg] & \multicolumn{1}{c} {[hours]} &  &  & \multicolumn{2}{c} {[hours]} & [km] \\
\hline\hline
\endhead
\hline
\endfoot
\hline
59     &  Elpis         &  300  &$-$55  &        &        &  13.6747  &  CE  &  289  &  13.67    &  3   &  165.1  \\
70     &  Panopaea      &  40   &  26   &  233   &  24    &  15.8041  &  C   &  314  &  15.81    &  3   &  127.9  \\
74     &  Galatea       &  109  &  1    &  289   &  14    &  17.2675  &  C   &  327  &  17.27    &  3   &  113.1  \\
83     &  Beatrix       &  182  &$-$9   &        &        &  10.1107  &  E   &  351  &  10.16    &  3   &  89.6   \\
105    &  Artemis       &  47   &  24   &        &        &  37.118   &  C   &  405  &  37.15    &  3   &  119.0  \\
106    &  Dione         &  57   &  12   &  237   &  1     &  16.2126  &  C   &  379  &  16.21    &  3   &  153.4  \\
126    &  Velleda       &  120  &  46   &  300   &  32    &  5.36708  &  C   &  249  &  5.37     &  3   &  43.9   \\
131    &  Vala          &  58   &  34   &  244   &  17    &  5.18072  &  C   &  368  &  5.18     &  3   &  31.3   \\
\end{longtable}
\tablefoot{
We provide ecliptic coordinates $\lambda$ and $\beta$ of up to two pole solutions, sidereal rotational period $P$ (its uncertainty is on the order of the last digit). The method column (M)  indicates whether the rotation period was derived by the convex inversion approach (C), the ellipsoid approach (E), or both (CE). The final columns are the number of ASAS-SN measurements $N_{\mathrm{A}}$, the LCDB period with its reliability flag \citep{Warner2009}, and the size $D$ from the MP3C database\footnote{\url{https://mp3c.oca.eu/catalogue/index.htm}}.
}

\onecolumn
\begin{landscape}
\begin{longtable}{r@{\,\,\,}l rrrr D{.}{.}{6} rrD{.}{.}{2}rrrrrrD{.}{.}{6}}
\caption{\label{tab:models_damit}Physical properties of asteroids with new shape model determinations based on the $V$ band ASAS-SN data.}\\
\hline 
\multicolumn{2}{c} {Asteroid} & 
\multicolumn{1}{c} {$\lambda_1$} &
\multicolumn{1}{c} {$\beta_1$} &
\multicolumn{1}{c} {$\lambda_2$} &
\multicolumn{1}{c} {$\beta_2$} &
\multicolumn{1}{c} {$P$} & M &
$N_{\mathrm{A}}$ & \multicolumn{2}{c}{$P_{\mathrm{LCDB}}$} & \multicolumn{1}{c} {$D$} & 
$\lambda_\mathrm{D1}$ & $\beta_\mathrm{D1}$ & 
$\lambda_\mathrm{D2}$ & $\beta_\mathrm{D2}$ & 
\multicolumn{1}{c}{$P_\text{D}$} \\
\multicolumn{2}{l} { } & [deg] & [deg] & [deg] & [deg] & \multicolumn{1}{c} {[hours]} &  &  & \multicolumn{2}{c} {[hours]} & [km] & [deg] & [deg] & [deg] & [deg] & \multicolumn{1}{c}{[hours]} \\
\hline\hline

\endfirsthead
\caption{continued.}\\

\hline
\multicolumn{2}{c} {Asteroid} &
\multicolumn{1}{c} {$\lambda_1$} &
\multicolumn{1}{c} {$\beta_1$} &
\multicolumn{1}{c} {$\lambda_2$} &
\multicolumn{1}{c} {$\beta_2$} &
\multicolumn{1}{c} {$P$} &M &
$N_{\mathrm{A}}$ & \multicolumn{2}{c} {$P_{\mathrm{LCDB}}$} & \multicolumn{1}{c} {$D$} & $\lambda_{\mathrm{D1}}$ & $\beta_{\mathrm{D1}}$ &
$\lambda_{\mathrm{D2}}$ & $\beta_{\mathrm{D2}}$ &
\multicolumn{1}{c}{$P_{\mathrm{D}}$}\\
\multicolumn{2}{l} { } & [deg] & [deg] & [deg] & [deg] & \multicolumn{1}{c} {[hours]} &  &  & \multicolumn{2}{c} {[hours]} & [km] & [deg] & [deg] & [deg] & [deg] & \multicolumn{1}{c} {[hours]} \\
\hline\hline
\endhead
\hline
\endfoot
\hline
5      &  Astraea          &  84   &  52    &  270   &  44    &  16.8006  &  CE  &  312  &  16.81    &  3   &  108.3  &  124  &  39   &        &        &  16.8006  \\
12     &  Victoria         &  170  &  37    &        &        &  8.6604   &  C   &  347  &  8.66     &  3   &  115.1  &  178  &$-$28  &        &        &  8.6603   \\
13     &  Egeria           &  34   &  46    &        &        &  7.04664  &  C   &  262  &  7.04     &  3   &  227.0  &  54   &  34   &  233   &  6     &  7.04667  \\
16     &  Psyche           &  35   &$-$11   &  215   &  16    &  4.19597  &  C   &  377  &  4.20     &  3   &  207.2  &  36   &$-$10  &        &        &  4.19595  \\
17     &  Thetis           &  55   &  12    &  237   &  21    &  12.2660  &  CE  &  290  &  12.27    &  3   &  84.9   &  236  &  19   &        &        &  12.2660  \\
19     &  Fortuna          &  102  &  51    &  283   &  59    &  7.4432   &  CE  &  252  &  7.44     &  3   &  196.4  &  97   &  69   &        &        &  7.4432   \\
20     &  Massalia         &  122  &  46    &  302   &  47    &  8.0976   &  CE  &  312  &  8.10     &  3   &  131.6  &  179  &  39   &  360   &  40    &  8.0990   \\
21     &  Lutetia          &  52   &$-$3    &  232   &$-$0    &  8.1683   &  CE  &  300  &  8.17     &  3   &  108.4  &  52   &$-$6   &        &        &  8.1683   \\
\end{longtable}
\tablefoot{
We provide ecliptic coordinates $\lambda$ and $\beta$ of up to two pole solutions, sidereal rotational period $P$ (its uncertainty is of the order of the last digit). The method (M) column indicates whether the rotation period was derived by the convex inversion (C), ellipsoid (E) approach, or both (CE) approaches. The final columns are the number of ASAS-SN measurements $N_{\mathrm{A}}$, the LCDB period with its reliability flag \citep{Warner2009}, the size $D$ from the MP3C database\footnote{\url{https://mp3c.oca.eu/catalogue/index.htm}}, and the DAMIT rotation state solution.
}
\end{landscape}
\onecolumn

\begin{longtable}{r@{\,\,\,}l D{.}{.}{6} rrr}
\caption{\label{tab:periods_new}Physical properties of asteroids with new rotation period determinations based on the $V$-band ASAS-SN data.}\\
\hline 
\multicolumn{2}{c} {Asteroid} & \multicolumn{1}{c} {$P$} & M & $N_{\mathrm{A}}$ & \multicolumn{1}{c} {$D$}\\
\multicolumn{2}{l} { } & \multicolumn{1}{c} {[hours]} &  &  & [km] \\
\hline\hline

\endfirsthead
\caption{continued.}\\

\hline
\multicolumn{2}{c} {Asteroid} & \multicolumn{1}{c} {$P$} & M & $N_{\mathrm{A}}$ & \multicolumn{1}{c} {$D$}\\
\multicolumn{2}{l} { } & \multicolumn{1}{c} {[hours]} &  &  & [km] \\
\hline\hline
\endhead
\hline
\endfoot
\hline

835    &  Olivia        &  1104.0   &  C   &  145  &  30.4  \\
991    &  McDonalda     &  339.5    &  E   &  178  &  38.6  \\
1034   &  Mozartia      &  449.9    &  E   &  216  &  9.7   \\
1445   &  Konkolya      &  59.395   &  C   &  96   &  20.3  \\
1714   &  Sy            &  317.8    &  E   &  193  &  14.0  \\
1745   &  Ferguson      &  1078.2   &  C   &  136  &  12.1  \\
1755   &  Lorbach       &  7.9765   &  CE  &  208  &  24.9  \\
1787   &  Chiny         &  13.5609  &  E   &  173  &  19.9  \\
1984   &  Fedynskij     &  8.1405   &  E   &  187  &  38.4  \\
2027   &  Shen Guo      &  1117.3   &  E   &  164  &  16.5  \\
2051   &  Chang         &  12.0142  &  CE  &  183  &  16.4  \\
2138   &  Swissair      &  369.7    &  C   &  244  &  12.9  \\
2158   &  Tietjen       &  8.6581   &  E   &  137  &  22.7  \\
2165   &  Young         &  6.39009  &  CE  &  143  &  27.1  \\
2191   &  Uppsala       &  31.317   &  E   &  213  &  17.5  \\
2219   &  Mannucci      &  27.460   &  E   &  172  &  39.1  \\
2226   &  Cunitza       &  189.05   &  E   &  159  &  14.6  \\
2248   &  Kanda         &  24.736   &  E   &  169  &  26.4  \\
2404   &  Antarctica    &  2.26238  &  CE  &  143  &  23.2  \\
2405   &  Welch         &  14.3411  &  E   &  123  &  26.4  \\
2413   &  van de Hulst  &  211.96   &  E   &  190  &  20.8  \\
2475   &  Semenov       &  12.2370  &  E   &  131  &  14.5  \\
2502   &  Nummela       &  14.3151  &  E   &  97   &  19.0  \\
2660   &  Wasserman     &  628.3    &  E   &  137  &  9.2   \\
2780   &  Monnig        &  681.8    &  E   &  91   &  4.8   \\
2795   &  Lepage        &  40.368   &  E   &  107  &  6.1   \\
2819   &  Ensor         &  161.66   &  E   &  130  &  9.7   \\
2837   &  Griboedov     &  3.94988  &  C   &  119  &  12.3  \\
2909   &  Hoshi-no-ie   &  311.3    &  CE  &  188  &  21.3  \\
2949   &  Kaverznev     &  83.81    &  CE  &  77   &  7.0   \\
2967   &  Vladisvyat    &  8.3739   &  E   &  176  &  32.9  \\
3158   &  Anga          &  10.3203  &  E   &  113  &  7.3   \\
3181   &  Ahnert        &  1187.1   &  E   &  134  &  8.0   \\
3214   &  Makarenko     &  249.2    &  E   &  165  &  21.2  \\
3311   &  Podobed       &  191.43   &  E   &  127  &  17.3  \\
3346   &  Gerla         &  26.239   &  E   &  208  &  34.6  \\
3365   &  Recogne       &  254.2    &  E   &  114  &  17.7  \\
3380   &  Awaji         &  87.74    &  E   &  125  &  $-$   \\
3650   &  Kunming       &  12.3703  &  E   &  111  &  26.6  \\
3681   &  Boyan         &  966.3    &  C   &  107  &  4.2   \\
3702   &  Trubetskaya   &  8.2905   &  E   &  207  &  18.2  \\
3863   &  Gilyarovskij  &  12.9588  &  E   &  114  &  6.5   \\
3955   &  Bruckner      &  7.5492   &  E   &  205  &  18.1  \\
4101   &  Ruikou        &  3.13391  &  E   &  76   &  8.2   \\ 
4103   &  Chahine       &  105.08   &  CE  &  208  &  13.3  \\
4121   &  Carlin        &  97.78    &  E   &  74   &  6.9   \\
4129   &  Richelen      &  288.8    &  E   &  75   &  5.4   \\
4256   &  Kagamigawa    &  1014.7   &  E   &  110  &  7.5   \\
4292   &  Aoba          &  291.1    &  C   &  97   &  24.7  \\
4380   &  Geyer         &  7.7282   &  E   &  120  &  16.9  \\
4593   &  Reipurth      &  5.9985   &  E   &  96   &  12.3  \\
5042   &  Colpa         &  169.62   &  E   &  151  &  20.2  \\
5077   &  Favaloro      &  4.59045  &  E   &  68   &  4.5   \\
5203   &  Pavarotti     &  6.93199  &  E   &  97   &  5.0   \\ 
5238   &  Naozane       &  80.61    &  E   &  69   &  5.9   \\
5650   &  Mochihito-o   &  601.3    &  CE  &  196  &  11.4  \\ 
5886   &  Rutger        &  3.32503  &  E   &  142  &  17.3  \\
6064   &  Holasovice    &  337.7    &  C   &  71   &  3.8   \\ 
6917   & $-$            &  351.9    &  CE  &  74   &  4.4   \\ 
7387   &  Malbil        &  7.5498   &  E   &  64   &  $-$   \\
10314  & $-$            &  1311.5   &  E   &  73   &  21.8  \\
10426  &  Charlierouse  &  36.975   &  E   &  147  &  9.2   \\
11358  & $-$            &  10.7927  &  E   &  84   &  13.6  \\
11923  & $-$            &  7.1288   &  E   &  62   &  2.8   \\ 
13249  &  Marcallen     &  190.9    &  E   &  123  &  20.8  \\
13832  & $-$            &  98.43    &  E   &  169  &  37.5  \\
14720  & $-$            &  108.14   &  E   &  149  &  8.5   \\
20602  & $-$            &  7.3061   &  E   &  79   &  23.7  \\
27396  &  Shuji         &  203.68   &  E   &  106  &  23.9  \\

\end{longtable}
\tablefoot{
We provide the sidereal rotational period $P$ (its uncertainty is on the order of the last digit). The method column (M)  indicates whether the rotation period was derived by the convex inversion approach (C), the ellipsoid approach (E), or both (CE). The final two columns are the number of ASAS-SN measurements $N_{\mathrm{A}}$ and the size $D$ from the MP3C database. 
}

\end{appendix}

\end{document}